\documentclass{article}

\usepackage[preprint,nonatbib]{neurips_2022}

\usepackage[citestyle=numeric-comp,sorting=none,maxbibnames=99,url=false,doi=false,giveninits=true]{biblatex}
\DeclareFieldInputHandler{title}{}
\usepackage{amsmath}
\usepackage{textcomp}  
\usepackage[utf8]{inputenc} 
\usepackage{amsmath}
\usepackage{booktabs}
\usepackage{float}
\usepackage{siunitx}
\usepackage{multirow}
\usepackage{hyperref}
\usepackage[utf8]{inputenc} 
\usepackage[T1]{fontenc}    
\usepackage{hyperref}       
\usepackage{url}            
\usepackage{booktabs}       
\usepackage{amsfonts}       
\usepackage{nicefrac}       
\usepackage{microtype}      
\usepackage[table]{xcolor}         
\usepackage[pdftex]{graphicx} 
\usepackage{threeparttable} 
\usepackage{booktabs}       
\usepackage{array}
\usepackage{caption}
\usepackage{subcaption}
\usepackage{pifont}

\usepackage[longend]{algorithm2e}
\usepackage{textgreek}
\usepackage{amssymb}
\usepackage[version=4]{mhchem}
\usepackage{tcolorbox}

\usepackage{amsmath}
\usepackage{bm}

\title{What Is Next for LLMs? Next-Generation AI Computing Hardware Using Photonic Chips}

\author{
  \small Renjie Li \textsuperscript{$1, 5, \dagger$} 
  \And
  \small Wenjie Wei \textsuperscript{$2, \dagger$} 
  \And
  \small Qi Xin \textsuperscript{$1$} 
  \And
  \small Xiaoli Liu \textsuperscript{$2$} 
  \And
  \small Sixuan Mao \textsuperscript{$1$} \\
  \and
   \small Erik Ma \textsuperscript{$6$} 
   \: \:
   \small Zijian Chen \textsuperscript{$5$} 
   \: \:
  \small Malu Zhang \textsuperscript{$*,2$} \: \: 
  \small Haizhou Li \textsuperscript{$*,3,4$} \: \:
  \small Zhaoyu Zhang \textsuperscript{$*,1$} \\
  \small \texttt{Email: zhangzy@cuhk.edu.cn} \\
\tiny \textsuperscript{1} School of Science and Engineering, Guangdong Key Laboratory of Optoelectronic Materials and Chips, \\ \tiny Shenzhen Key Lab of Semiconductor Lasers, The Chinese University of Hong Kong, Shenzhen \\
\tiny \textsuperscript{2} University of Electronic Science and Technology of China, \\
\tiny \textsuperscript{3} School of Data Science, The Chinese University of Hong Kong, Shenzhen\\
\tiny \textsuperscript{4}  National University of Singapore \\
\tiny \textsuperscript{5}  University of Illinois Urbana-Champaign \\
\tiny \textsuperscript{6}  University of California, Berkeley\\
\small \textsuperscript{*} indicates corresponding authors, \: \textsuperscript{$\dagger$} indicates equal contribution
}

\begin{document}

\maketitle
\begin{abstract}

Large language models (LLMs) are rapidly pushing the limits of contemporary computing hardware. For example, training GPT-3 has been estimated to consume around 1300 MWh of electricity, and projections suggest future models may require city-scale (gigawatt) power budgets. These demands motivate exploration of computing paradigms beyond conventional von Neumann architectures. This review surveys emerging photonic hardware optimized for next-generation generative AI computing. We discuss integrated photonic neural network architectures (e.g. Mach–Zehnder interferometer meshes, lasers, wavelength-multiplexed microring-resonators) that perform ultrafast matrix operations. We also examine promising alternative neuromorphic devices, including spiking neural network circuits and hybrid spintronic-photonic synapses, which combine memory and processing. The integration of two-dimensional materials (graphene, TMDCs) into silicon photonic platforms is reviewed for tunable modulators and on-chip synaptic elements. Transformer-based LLM architectures (self-attention and feed-forward layers) are analyzed in this context, identifying strategies and challenges for mapping dynamic matrix multiplications onto these novel hardware substrates. We then dissected the mechanisms of mainstream LLMs, such as chatGPT, DeepSeek, and Llama, highlighting their architectural similarities and differences. We synthesize state-of-the-art components, algorithms, and integration methods, highlighting key advances and open issues in scaling such systems to mega-sized LLM models. We find that photonic computing systems could potentially surpass electronic processors by orders of magnitude in throughput and energy efficiency, but require breakthroughs in memory especially for long-context windows and long token sequences and in storage of ultra-large datasets. This survey provides a comprehensive roadmap for AI hardware development, emphasizing the role of cutting-edge photonic components and technologies in supporting future LLMs.

\end{abstract}

\newpage
\tableofcontents
\newpage

\section{Introduction}
The recent proliferation of transformer-based large language models (LLMs) has dramatically increased the demands on computing infrastructure. Training state-of-the-art AI models now requires enormous compute and energy resources. For example, the GPT-3 model consumed an estimated $1.3\times10^3$ MWh of electricity during training, and industry projections suggest that next-generation LLMs may demand power budgets on the order of gigawatts. These trends coincide with the use of massive GPU clusters (for instance, Meta has trained Llama 4 on a cluster exceeding $10^5$ NVIDIA H100 GPUs). Meanwhile, conventional silicon scaling is approaching fundamental limits (transistors are reaching $\sim$3 nm feature sizes), and von Neumann architectures suffer from memory-processor bottlenecks that constrain speed and energy efficiency \cite{li2025photonics}. Together, these factors underscore a growing gap between the computational demands of LLMs and the capabilities of traditional CMOS electronic hardware \cite{li2025photonics}. These challenges have spurred exploration of alternative computing paradigms. Photonic computing, which processes information with light, offers intrinsic high bandwidth, massive parallelism, and minimal heat dissipation. Recent advances in photonic integrated circuits (PICs) have enabled neural-network primitives such as coherent interferometer meshes, microring-resonator (MRR) weight banks, and wavelength-division multiplexing (WDM) schemes to perform dense matrix multiplications and multiply-accumulate operations at the speed of light. Such photonic processors exploit WDM to achieve extreme parallelism and throughput. Simultaneously, integrating two-dimensional (2D) materials (graphene, TMDCs) into PICs has produced ultrafast electro-absorption modulators and saturable absorbers that serve as on-chip neurons and synapses. Complementary to optics, spintronic neuromorphic devices (e.g., magnetic tunnel junctions and skyrmion channels) offer non-volatile synaptic memory and spiking neuron behavior. These photonic and spintronic neuromorphic elements inherently co-locate memory and processing and leverage new physical mechanisms for energy-efficient AI computation. Mapping transformer-based LLM architectures onto these emerging hardware substrates raises unique challenges. Transformer self-attention layers involve dynamically computed weight matrices (queries, keys, and values) that depend on the input data. Designing reconfigurable photonic or spintronic circuits to realize such data-dependent operations is an active area of research. Furthermore, implementing analog nonlinearities (e.g. GeLU activation) and normalization in optical/spintronic media remains a major challenge. Addressing these issues has motivated hardware-aware algorithm design, such as photonics-friendly training methods and neural network models that tolerate analog noise and quantization.

The remainder of this review is organized as follows. Section 2 surveys photonic accelerator architectures, including coherent interferometer meshes, microring weight banks, and WDM-based matrix processors. Section 3 discusses integration of two-dimensional materials into photonic chips (graphene/TMDC modulators, photonic memristors). Section 4 examines alternative neuromorphic devices, covering spintronics for neuromorphic computing.  Section 5 summarizes the principles of mainstream LLMs and transformers and how they can be mapped onto photonic chips, highlighting strategies for implementing attention and feed-forward layers in photonic and neuromorphic hardware. Section 6 introduces the mechanisms and algorithms of spiking neural networks and their implementation.  Finally, Section 7 identifies key system-level challenges and outlines future directions. Through this comprehensive survey, we aim to chart a roadmap for next-generation AI hardware development using photonic and spintronic technologies.

\section{State-of-The-Art Photonic Components for Photonic Neural Networks and Photonic Computing}

Photonic neural networks (PNN) leverage the synergistic effects of various optical components to achieve efficient computation: microring resonators utilize resonance effects for wavelength multiplexing and optical frequency comb generation, providing the foundation for multi-wavelength signal processing \cite{xu202111, feldmann2021parallel, cheng2023human, xu202111, feldmann2021parallel, cheng2023human}; Mach-Zehnder interferometer (MZI) arrays perform optical matrix operations through phase modulation, enabling core linear transformations in neural networks \cite{shen2017deep, hughes2018training, zhu2022space, pai2023experimentally}; metasurfaces manipulate the phase and amplitude of light waves via subwavelength structures, executing highly parallel optical computations in the diffraction domain \cite{lin2018all, wang2019chip, qian2020performing, zarei2020integrated, goi2021nanoprinted, zhou2021large, luo2022metasurface, liu2022programmable, fu2023photonic}; the 4f system performs linear filtering in the diffraction domain through Fourier transform \cite{yan2019fourier, chang2018hybrid}; while novel lasers achieve nonlinear activation through electro-optic conversion via diverse approaches \cite{xiang2020computing, chen2023deep, shi2023photonic, xiang2023hardware}. By integrating optical field manipulation, linear transformations, and nonlinear responses, these components construct all-optical computing architectures with high speed, low power consumption, and massive parallelism. This section introduces the optical devices commonly employed in current ONN implementations.

\subsection{Microring resonator}

\begin{figure}[hbtp]
\centering
\includegraphics[width=1.\textwidth]{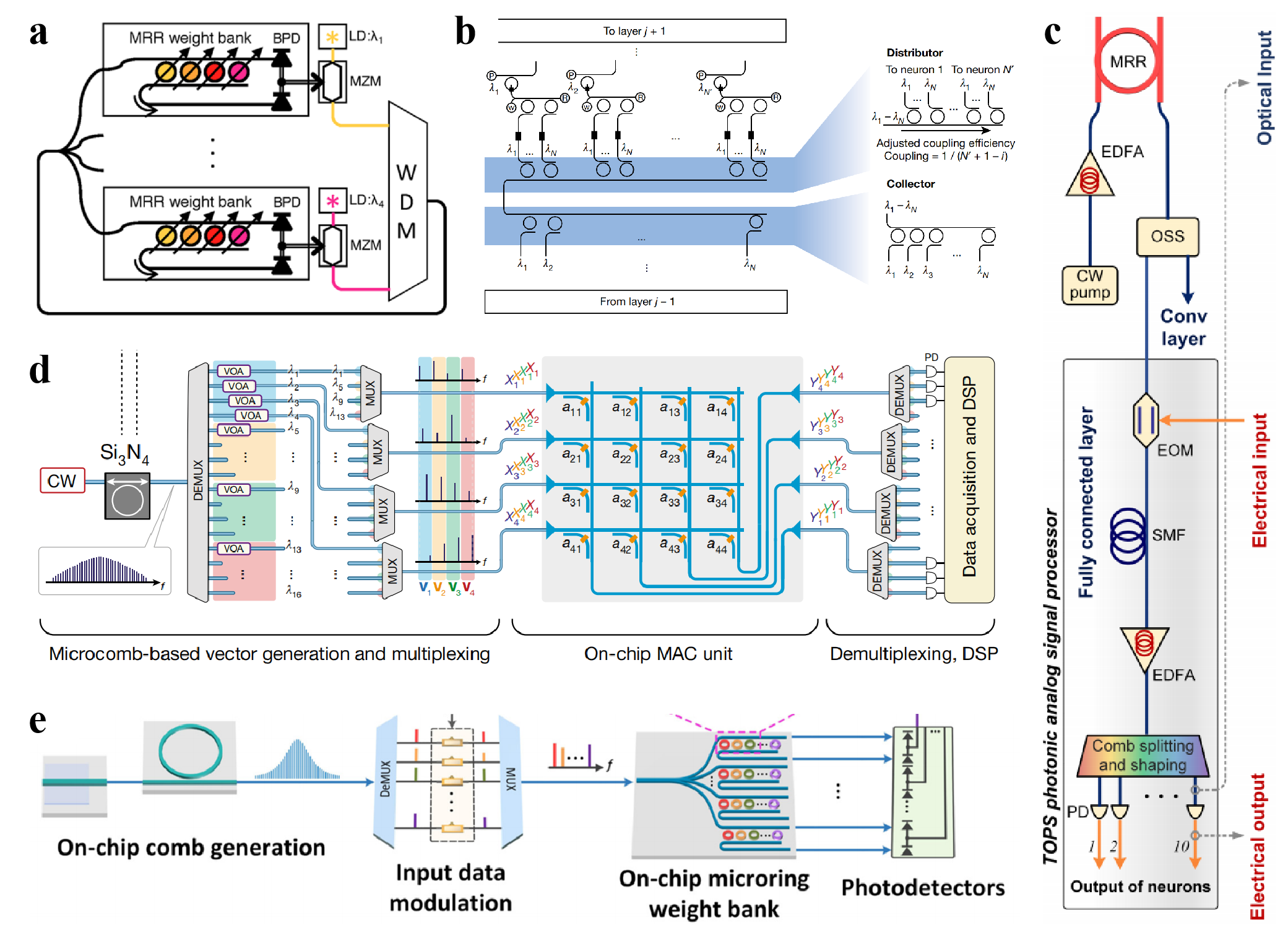}
\caption{Microring resonator: 
a Neuromorphic ONNs can be realized through microring resonator (MRR) weight banks. \cite{tait2017neuromorphic} 
b The underlying mechanism and experimental setup of fully optical spiking neural networks are illustrated in \cite{feldmann2019all}.
c A photonic convolution accelerator has been developed using a time-wavelength multiplexing approach. \cite{xu202111}
d In-memory photonic computing architectures leverage on-chip microcombs and phase-change materials. \cite{feldmann2021parallel}
e Microcomb-based integrated ONNs enable convolution operations for applications such as emotion recognition. \cite{cheng2023human}
}\label{fig:MRR}
\index{figures}
\end{figure}

The significance of microring resonators (MRRs) (Fig. \ref{fig:MRR}) extends beyond their role as waveguides for wavelength-division multiplexing (WDM) \cite{xu202111, feldmann2021parallel, cheng2023human} to their unique filtering capabilities, such as optical frequency comb generation \cite{xu202111, feldmann2021parallel, cheng2023human}. On the one hand, WDM allows simultaneous propagation of different wavelength signals in the same structure without inter-channel interference: By designing the radius and refractive index of MRRs to support specific resonant wavelengths, light matching the resonance condition becomes coupled into the ring cavity for sustained oscillation, manifesting as distinct absorption dips in the transmission spectrum. On the other hand, the optical frequency combs arise from parametric oscillations in high-Q (low-loss) microresonators: When a continuous-wave (CW) pump laser is injected, photons experience nonlinear effects (e.g., Kerr nonlinearity), spontaneously generating equidistant spectral lines that form a comb-like spectrum. The interplay between WDM and comb generation allows multi-wavelength signals to be simultaneously synthesized and transmitted through shared waveguides, achieving both wavelength multiplexing and spatial multiplexing capabilities.

Other properties of microrings have also been exploited. For example, paper \cite{feldmann2019all} utilized the thermo-optic effect of microrings. paper \cite{feldmann2019all, feldmann2021parallel} installed phase-change materials with a lasing threshold on the ring to achieve a nonlinear effect similar to the ReLU function in neural networks.

\subsection{Mach-Zehnder Interferometer}

\begin{figure}[hbtp]
\centering
\includegraphics[width=1.\textwidth]{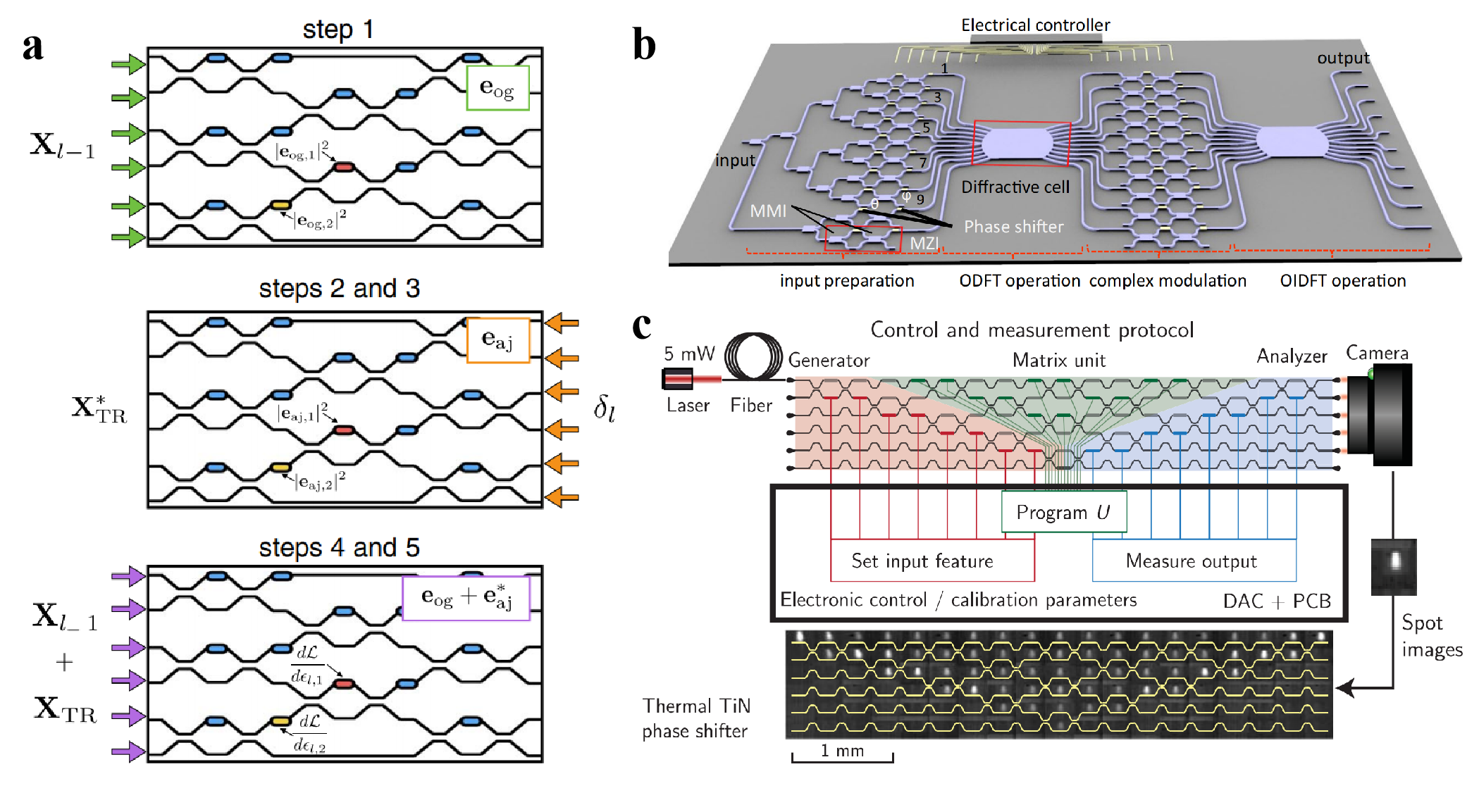}
\caption{Mach-Zehnder Interferometer: 
a Training methodology diagram for ONNs enabling real-time in-situ learning \cite{hughes2018training} 
b Integrated photonic neural network architecture combining MZIs with diffractive optical components \cite{zhu2022space}
c Demonstrated in situ backpropagation training of a photonic neural network using MZI meshes. \cite{pai2023experimentally}
}\label{fig:MZI}
\index{figures}
\end{figure}

MZI arrays (Fig. \ref{fig:MZI}) can effectively performing optical matrix-vector multiplication (MVM) \cite{shen2017deep, hughes2018training, zhu2022space, pai2023experimentally}: It is composed of two optical couplers/splitters and two modulators (which can be controlled via external circuits). The input light is split into two arms by the splitter, and the phase difference between them is adjusted by the modulators. Finally, the light is recombined through the optical coupler, resulting in interference. Each MZI performs a 2D unitary transformation (orthogonal transformation in the complex domain) on optical signals, mathematically equivalent to a $2 \times 2$ unitary matrix. When multiple MZIs are cascaded in specific topologies (e.g., mesh configurations), their collective behavior corresponds to the decomposition of a high-dimensional unitary matrix since any N-dimensional unitary matrix can be decomposed into a sequence of 2D unitary operations. Thus MZI arrays can implement programmable unitary transformations analogous to weight matrices in neural networks. 

The output optical signals can be further converted through optoelectronic means and integrated with electronic devices to implement nonlinear activation functions, completing the forward propagation of the neural network.

\subsection{Metasurface}

\begin{figure}[hbtp]
\centering
\includegraphics[width=1.\textwidth]{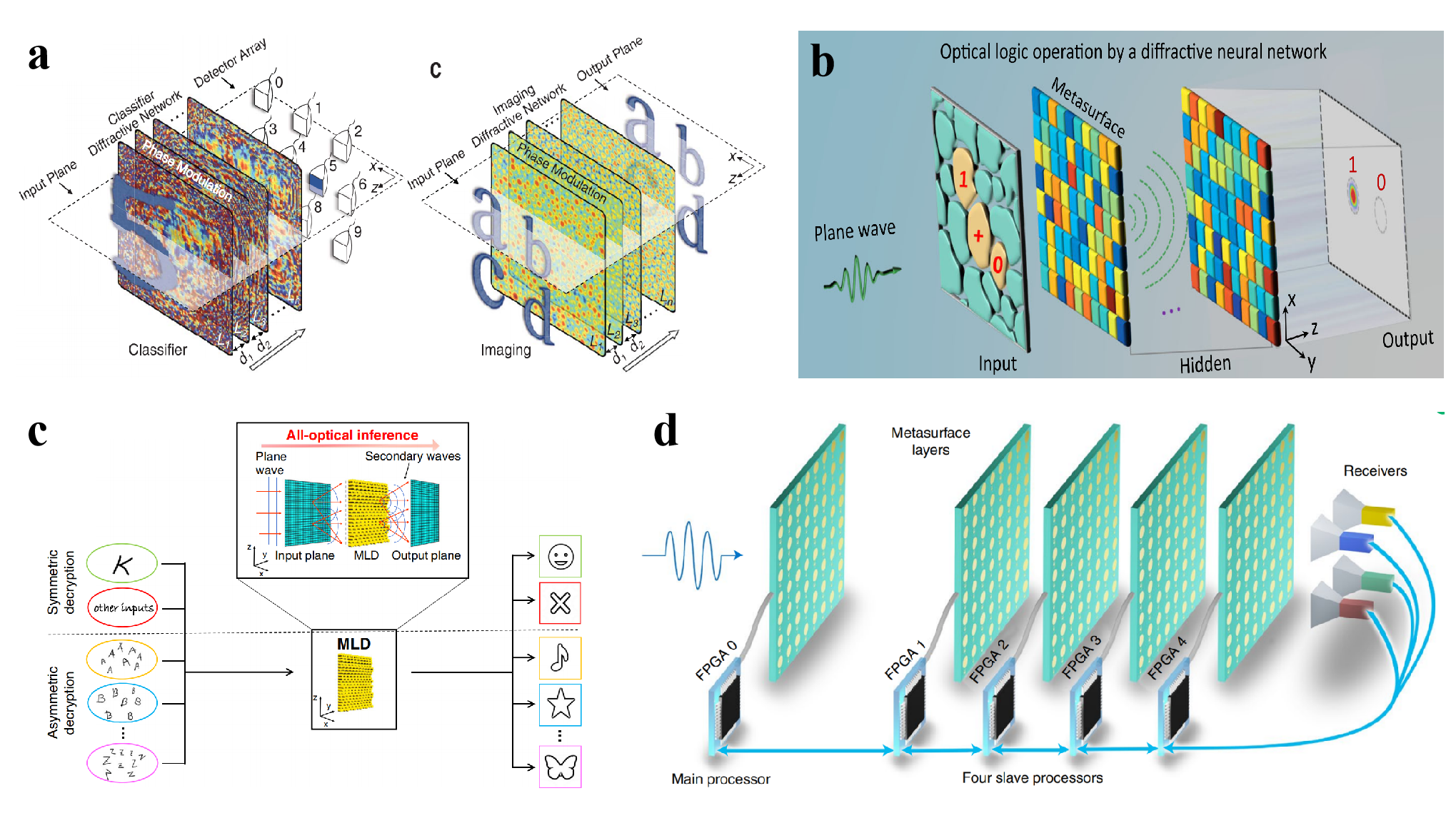}
\caption{2D Metasurface: 
a Conceptual representation of the inference mechanism in diffractive deep neural networks (D2NN). \cite{lin2018all}
b Experimental configuration demonstrating logical operations through diffractive optical neural networks (DONN). \cite{qian2020performing}
c Nanoprinted optical perceptrons enable on-chip. \cite{goi2021nanoprinted}
d Reconfigurable DONN architecture utilizing digital meta-atom arrays. \cite{liu2022programmable}
}\label{fig:meta-2D}
\index{figures}
\end{figure}

\begin{figure}[hbtp]
\centering
\includegraphics[width=0.8\textwidth]{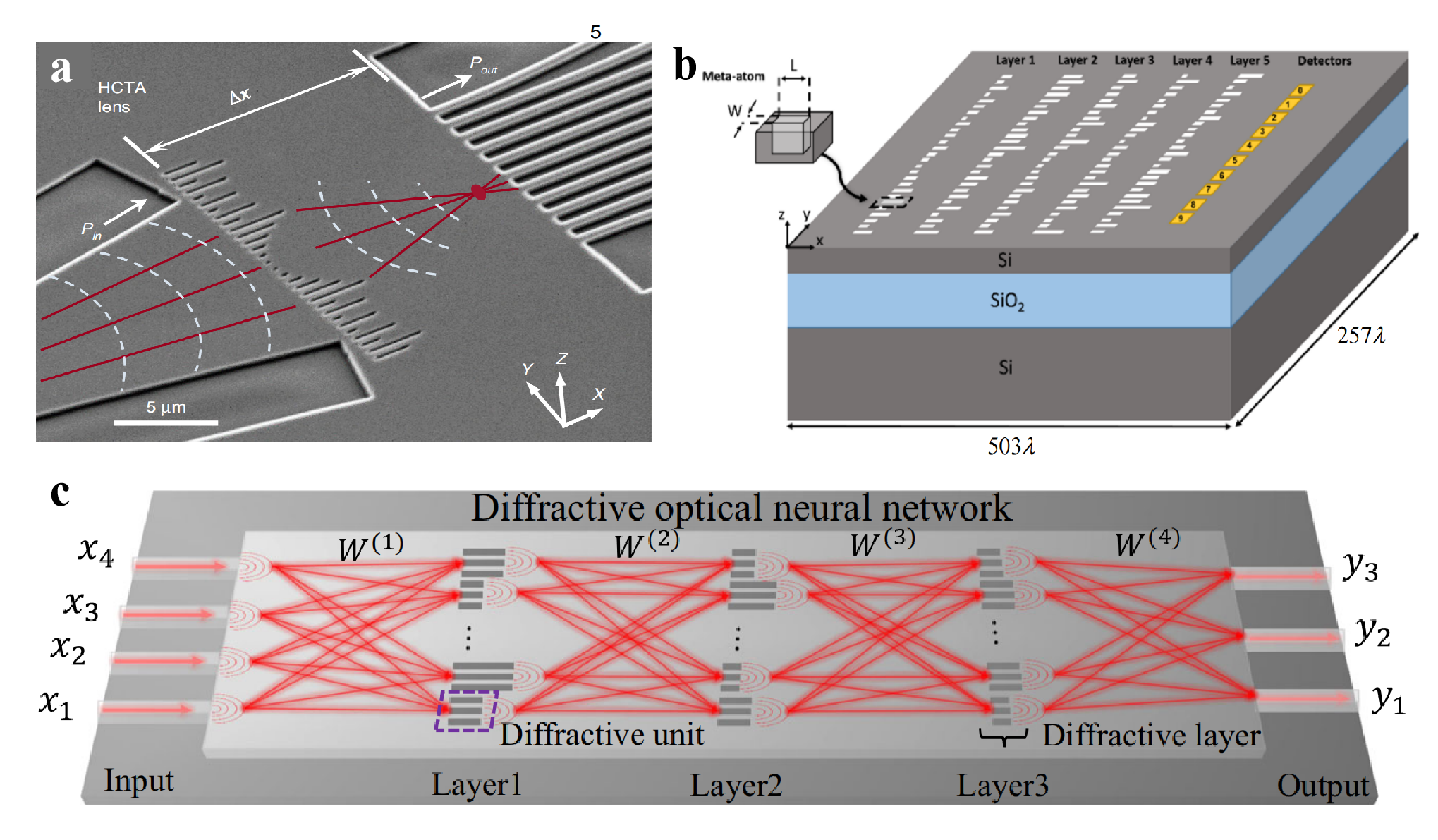}
\caption{1D Metasurface: 
a Experimental validation of 1D DONNs for photonic machine learning. \cite{wang2019chip}
b Simulation-based validation of on-chip DONN with light-speed computation. \cite{zarei2020integrated}
c Dielectric metasurface enables on-chip wavefront control for Fourier transform and spatial differentiation. \cite{fu2023photonic}
}\label{fig:meta-1D3}
\index{figures}
\end{figure}

The operation of metasurfaces in neural network applications primarily relies on the diffraction and interference of light between “surfaces”. \cite{lin2018all, wang2019chip, qian2020performing, zarei2020integrated, goi2021nanoprinted, zhou2021large, luo2022metasurface, liu2022programmable, fu2023photonic}. A metasurface is a material composed of subwavelength-scale structural elements that can modulate optical wave properties including phase, amplitude, polarization, and frequency. These structures typically exhibit ultra-thin profiles, lightweight characteristics, and high integration density (with massive parallelism), with diverse implementations such as silicon-on-insulator (SOI)-based designs \cite{wang2019chip, zarei2020integrated}, compound Huygens' metasurfaces \cite{qian2020performing}, and single-layer holographic perceptrons \cite{goi2021nanoprinted}. Since diffraction and interference are inherently linear processes, achieving nonlinear computation requires additional mechanisms, such as leveraging the optoelectronic effects of metasurface materials \cite{zhou2021large}.

Multilayer diffractive architectures (Fig. \ref{fig:meta-2D}) \cite{lin2018all, qian2020performing, zhou2021large, luo2022metasurface, liu2022programmable} employ stacked 2D surfaces as densely arranged neuron layers. Through controlled modulation of relative thickness or material properties at each spatial position in the diffraction layers, phase and amplitude adjustments of light are achieved. Alternatively, \cite{wang2019chip, zarei2020integrated, fu2023photonic} fabricate 1D high-contrast transmit array metasurfaces (Fig. \ref{fig:meta-1D3}) on one planar surface. For example, etching air grooves (potentially later filled with silica) on standard SOI substrates, featuring fixed groove spacing (lattice constants) and width. Phase control is achieved by modulating groove length. 

\textbf{4f system}

\begin{figure}[hbtp]
\centering
\includegraphics[width=1.\textwidth]{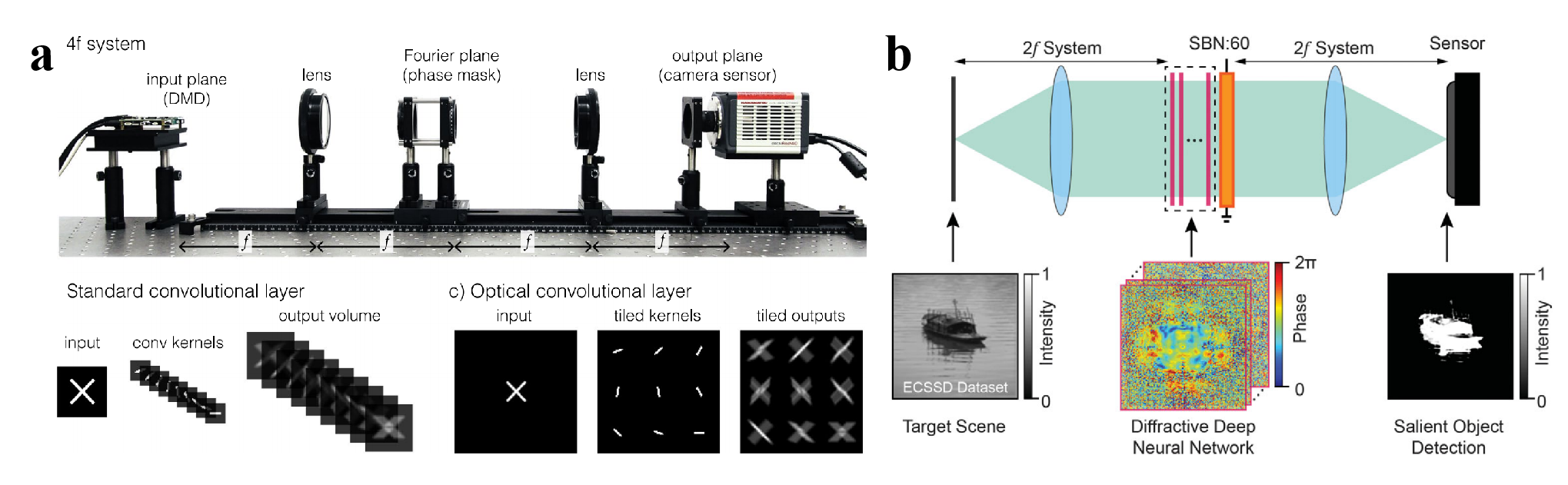}
\caption{4f system: 
a A hybrid optoelectronic CNN using 4f optical setup. \cite{chang2018hybrid}
b An entirely ONN architecture where a deep diffractive neural network is integrated into the Fourier plane of a 4f imaging system. \cite{yan2019fourier}
}\label{fig:4f}
\index{figures}
\end{figure}

The 4f system (Fig. \ref{fig:4f}) \cite{chang2018hybrid} employs optical field signals (e.g., images) that undergo Fourier transformation through the first lens. At the Fourier plane behind the lens, modulation devices (such as phase masks, spatial light modulators SLMs) perform spectral filtering or weight adjustment. The modulated spectrum is then inversely Fourier-transformed by a second lens to generate the output optical field. Metasurface materials may substitute traditional modulation devices between lenses \cite{yan2019fourier}.

\subsection{Other types of laser}

\begin{figure}[hbtp]
\centering
\includegraphics[width=1.\textwidth]{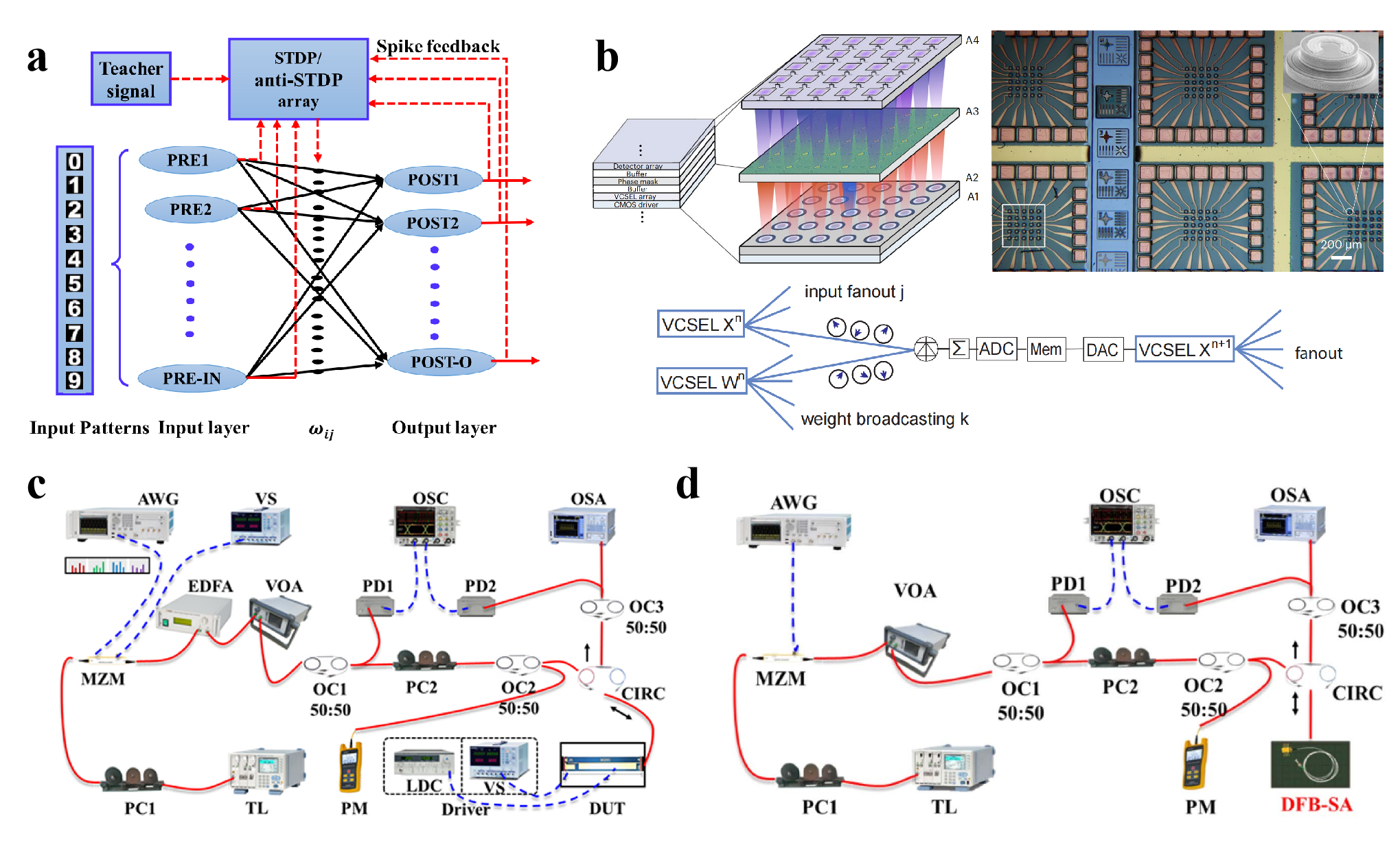}
\caption{Other types of laser: 
a  Theoretical analysis of the all-optical SNN using VCSELs. \cite{xiang2020computing}
b  VCSEL-based all-optical SNN for supervised learning. \cite{chen2023deep}
c  FP-SA neuron chip for hardware-algorithm collaborative computing in SNN \cite{xiang2023hardware}
d  Experimental demonstration of a photonic integrated spiking neuron using a DFB-SA laser. \cite{shi2023photonic}
}\label{fig:laser}
\index{figures}
\end{figure}

Lasers, as a unique light source characterized by high coherence, monochromaticity, and directionality, are also utilized in ONNs (Fig. \ref{fig:laser}). 

For instance, Vertical-Cavity Surface-Emitting Lasers (VCSELs) have been theoretically proposed and experimentally demonstrated in studies \cite{xiang2020computing, chen2023deep}. In a VCSEL, current is injected through the electrodes into the active region, where electrons and holes recombine in the quantum well layers, emitting photons. These photons are reflected back and forth between two Distributed Bragg Reflector (DBR) mirrors, passing through the active region repeatedly and being amplified. When the gain (light amplification capability) exceeds the cavity losses (absorption, scattering, etc.), the threshold condition is met, and laser output is achieved \cite{chen2023deep}. One study leveraged the property of VCSEL arrays, which can maintain the same initial phase when mode-locked by a Leader Laser. In this work, feature data was encoded into electrical signals to modulate the pump voltage of one VCSEL, thereby adjusting its output light phase. Similarly, each column of the weight matrix was encoded into electrical signals to adjust the output light phases of other VCSELs. Beam splitters and couplers were used to allow the output light from the VCSEL corresponding to MNIST data to interfere with the output light from other VCSELs. Photodetectors collected the optical signals, which were summed into electrical signals as the input for the next layer of the VCSEL array, enabling forward propagation. In the final layer, the photodetector with the strongest output electrical signal corresponded to the output label.

Another example is the Distributed Feedback (DFB) laser with an intracavity saturable absorber (SA), referred to as DFB-SA \cite{shi2023photonic}. The DFB laser's cavity incorporates a periodic grating structure, providing optical feedback to achieve single-wavelength output. The saturable absorber (SA) region is located near the high-reflectivity end of the laser cavity. At low pump levels, the SA absorbs photons, suppressing laser output; at high pump levels, the SA allows the release of optical pulses (Q-switching effect). Therefore, when the gain current exceeds the self-pulsation threshold of the DFB-SA, the periodic absorption modulation of the SA results in pulsating output, and the output frequency exhibits a nonlinear positive correlation with the pump intensity, which can serve as the fundamental unit of a Spiking Neural Network (SNN). Here, the DFB laser can also be replaced by a traditional Fabry-Perot (FP) laser \cite{xiang2023hardware}.

\section{Using 2D Materials to Make Integrated Photonic Chips}

Another piece of technology that is emerging as a critical component for next-generation AI hardware are integrated photonic chips, which leverage light for computation and communication allowing for both high speed and energy efficiency. For such an application, the integration of two-dimensional (2D) materials, mainly graphene and transition metal dichalcogenides (TMDCs), offers unique advantages improving both functionality and performance for today's photonic chips. This section will be exploring the properties, integration techniques, applications, as well as current challenges associated with the use of these materials in photonic chips intended for AI workloads. 

\subsection{Key Properties of Graphene and TMDCs}

\begin{figure}[H]
\centering
\includegraphics[width=1.\textwidth]{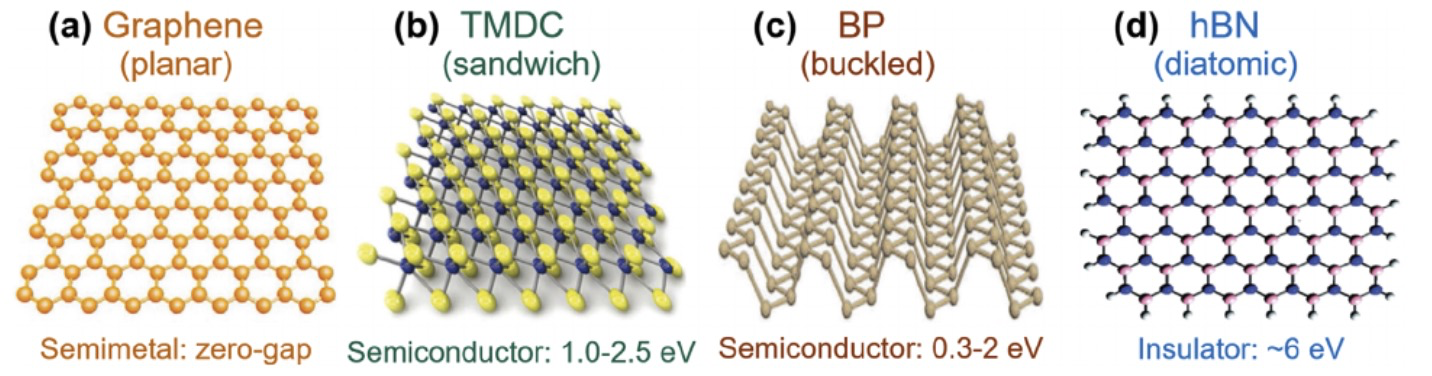}
\caption{Crystal structures of 
a) graphene,
b) TMDC,
c) black phosphorus,
d) hexagonal boron nitride
\cite{You2020Hybrid}
}\label{fig:2DMatCrystal}
\index{figures}
\end{figure}

As aforementioned, graphene and TMDCs are both promising materials being heavily researched for application in integrated photonic chips, and they both exhibit a range of properties that make them particularly suitable for such contexts. The former, graphene, has revolutionized the field of photonics because of its exceptional optical and electronic properties, being able to absorb approximately 2.3\% of incident light across a wide spectral range despite its minuscule thickness of only one atom. This is notable as it makes it highly effective for optical modulation and detection purposes \cite{xia2009ultrafast}. At the same time, this ultrafast carrier mobility allows for high-speed modulation and low-power operation, both qualities extremely critical for energy-efficient AI hardware \cite{li2014ultrafast, Cheng2021two}. Additionally, graphene exhibits strong nonlinear optical properties, something that is essential for frequency conversion, all-optical switching, and many other advanced functions, furthering its significance as a material in this given context. 

On the other hand, TMDCs, such as MoS$_2$ and WS$_2$, complement graphene by offering tunable bandgaps and strong excitonic effects. These materials exhibit enhanced light-matter interaction due to their direct bandgap in monolayer form, making them ideal for applications in photodetectors and waveguides \cite{wu2020two}. TMDCs also demonstrate strong nonlinear optical responses, enabling advanced functionalities like harmonic generation and parametric amplification on-chip \cite{Busschaert2020transition}. 

With these detailed characteristics and natural advantages posed by the utilization of graphene and TMDCs, it is clear why these two materials are crucial to the advancement of AI hardware in the realm of integrated photonic chips. 

\subsection{Integration Techniques}

The integration of 2D materials into photonic chips involves several advanced packaging techniques, which are detailed as follows: 

\textbf{Transfer Printing} Thin layers of 2D materials are exfoliated and transferred onto silicon substrates without adhesives, preserving intrinsic optical properties of the materials while allowing for precise placement onto photonic structures (waveguides, resonators, etc.) \cite{Cheng2021two}.

\begin{figure}[h]
\centering
\includegraphics[width=0.5\textwidth]{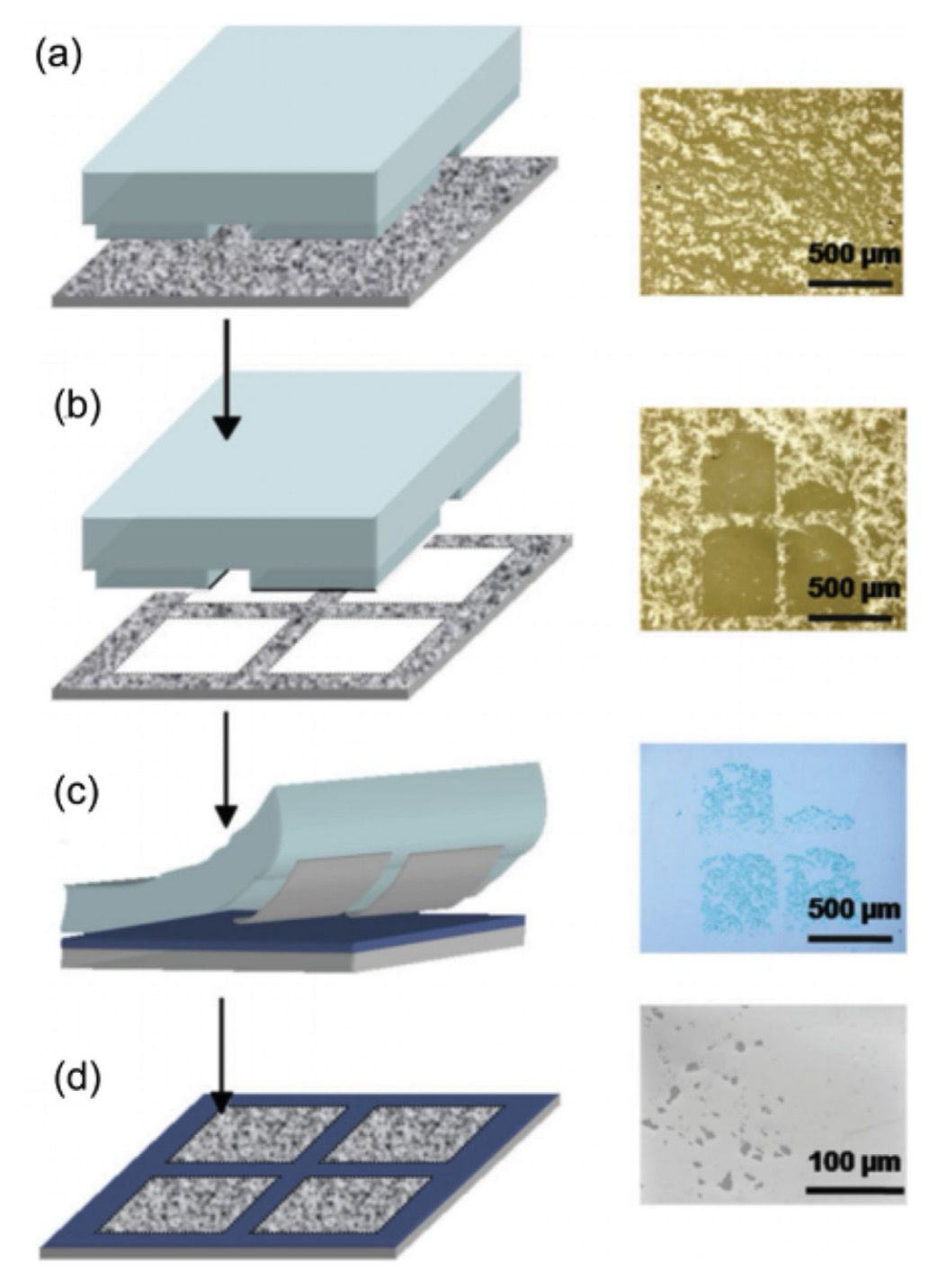}
\caption{Diagram (left) and images aquired from optical microscope (right) shocasing the softy exfoliation and transfer method, one of the main mechanical  methods of today. The process follows
a) depositing materials on glass substrate,
b) ink the pre-patterned polydimethylsiloxane (PDMS) stamp carefully,
c) contact inked stamp to heated Si/SiO2 substrate,
d) peel away revealing deposited materials.
\cite{Cao2021Recent}
}\label{fig:transferprint}
\index{figures}
\end{figure}

\textbf{Hybrid Integration} Combining graphene or TMDCs with existing silicon photonics platforms is another technique, one which enhances light-matter interaction. For example, graphene has been used to create high-speed modulators integrated into microring resonators. These hybrid devices achieve terahertz modulation speeds while maintaining low power consumption \cite{Marquez2020Graphene}.

\textbf{Van der Waals Heterostructures} For this technique, stacking different 2D materials enables the creation of heterostructures with tailored optical properties, such as tunable bandgaps and anisotropic refractive indices. These heterostructures are viewed to be useful for waveguiding applications where confinement factors need optimization \cite{Pelgrin2023Hybrid, You2020Hybrid, Cao2021Recent}.

\begin{figure}[h]
\centering
\includegraphics[width=1.\textwidth]{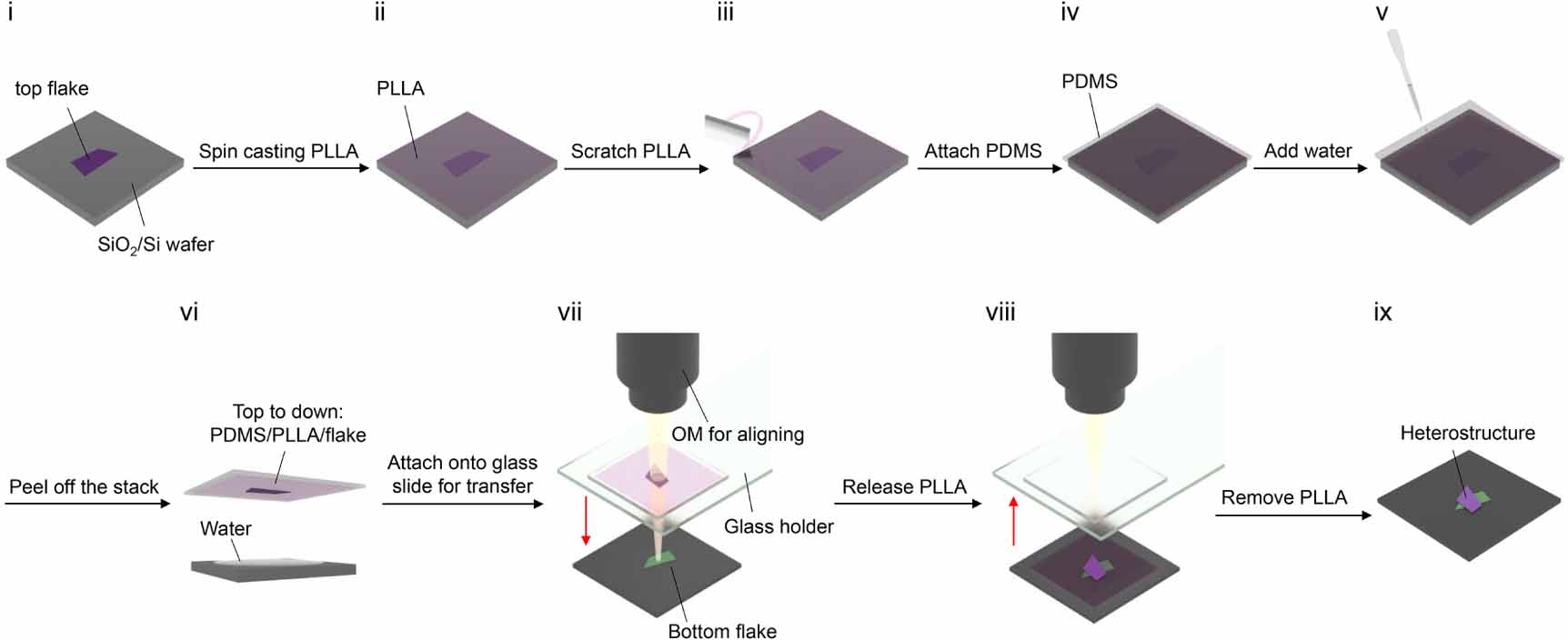}
\caption{Depiction of a schematic flow of the water immersion method used for constructing Van der Waals heterostructure without etchant. 
\cite{Cao2021Recent}
}\label{fig:vdwconstruct}
\index{figures}
\end{figure}

Recent advancements have also demonstrated the plausibility of utilizing wafer-scale integration of graphene-based devices using CMOS-compatible processes. This breakthrough paves the way for mass production of photonic chips incorporating 2D materials \cite{Pelgrin2023Hybrid}.

\subsection{Applications in Photonic Chips}

Photonic chips integrated with graphene and TMDCs offer transformative applications across AI workloads:

\textbf{Optical Modulators} Graphene-based modulators have demonstrated exceptional speed and bandwidth performance - by integrating graphene with silicon waveguides, researchers have achieved modulators capable of operating at frequencies exceeding 100 GHz \cite{Marquez2020Graphene}. These modulators are particularly suited for high-speed data transfer use cases found in AI systems.

\textbf{Photodetectors} A rather surprising application for graphene is in photodetectors, as the frequency-independent absorption nature coupled with extremely high carrier mobility when utilized with strong light absorbing materials created photodetectors outperforming the typical materials \cite{graphenea}. Research has taken an interesting direction in using hybrid graphene-quantum dot photodetectors as broadband image sensors inserted into CMOS cameras to achieve high responsivity \cite{graphenea}. Generally speaking, 2D materials have several advantages in terms waveguide-integrated photodetectors, including minimized dimensions, improved signal-to-noise ratio, and increased efficiency in broad bandwidth as well as high quantum applications \cite{wu2020two}. 

\begin{figure}[h]
\centering
\includegraphics[width=1.\textwidth]{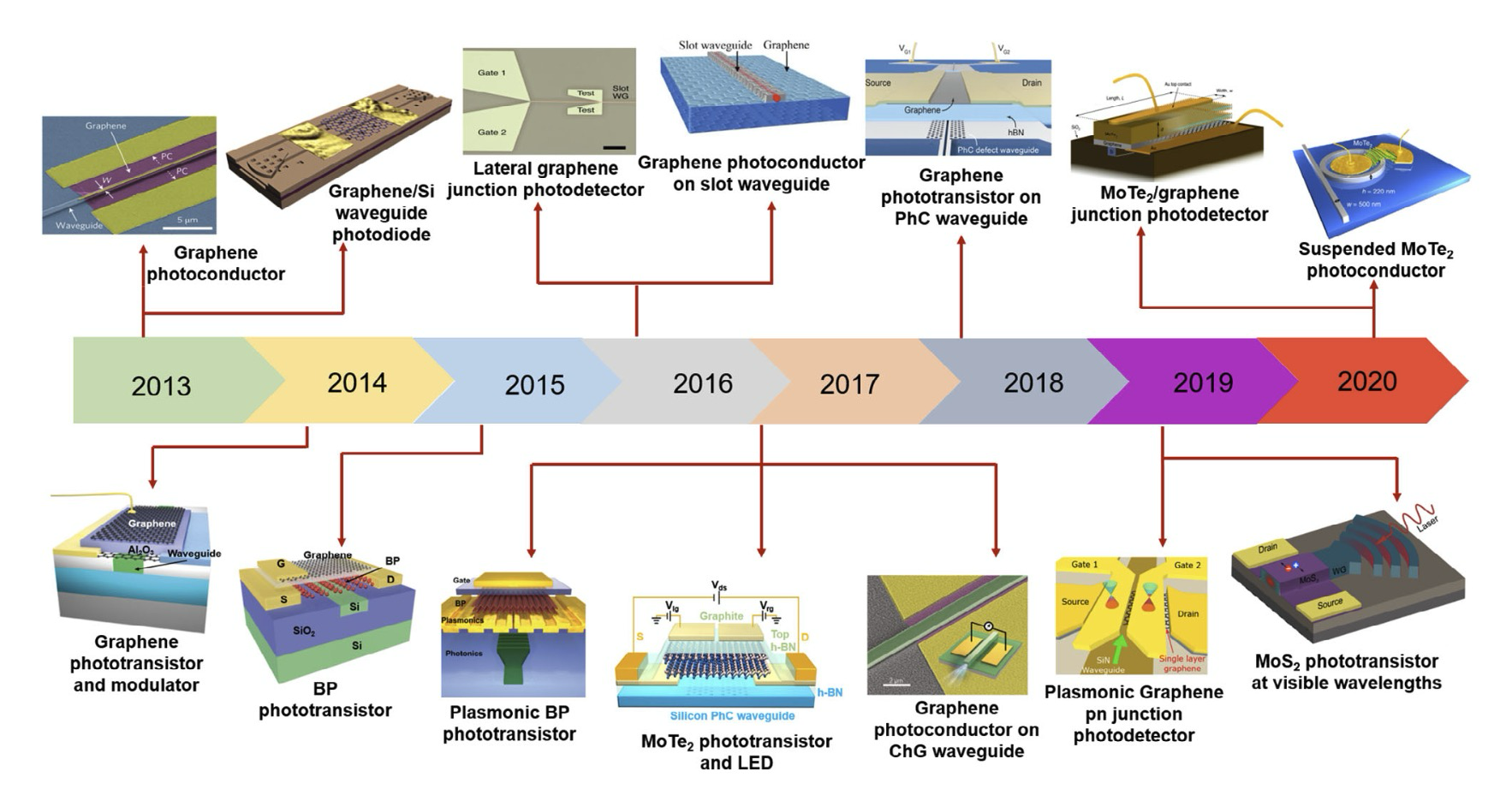}
\caption{Roadmap of waveguide-integrated photodetectors that are dependent on 2D materials.  
\cite{wu2020two}
}\label{fig:photodetectdev}
\index{figures}
\end{figure}

TMDCs are used to fabricate photodetectors with high responsivity across visible and infrared wavelengths, leveraging their physical properties to enhance such performances. Such detectors enable efficient data acquisition for AI-driven edge devices \cite{You2020Hybrid}. Hybrid graphene-quantum dot photodetectors have also been researched, striving to further enhance broadband detection capabilities while maintaining CMOS compatibility \cite{You2020Hybrid}.

\begin{figure}[H]
\centering
\includegraphics[width=0.65\textwidth]{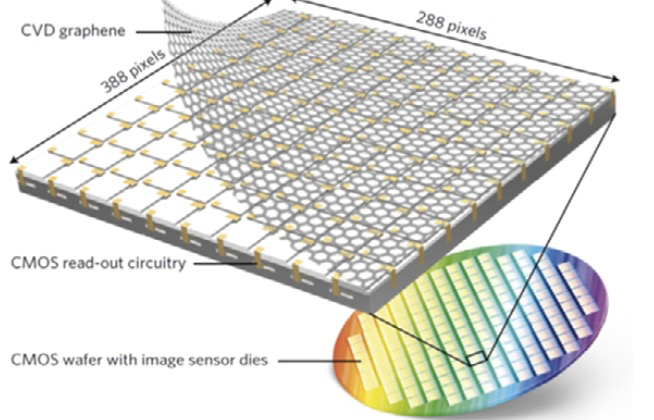}
\caption{Graphene-quantum dot photodetector implemented within CMOS circuits.  
\cite{You2020Hybrid}
}\label{fig:gqdphotodetect}
\index{figures}
\end{figure}

\textbf{Waveguides} The use of van der Waals materials has enabled ultrathin waveguides with low propagation losses. And through combining silicon photonics with waveguide-integrated graphene, properties such as full tunability, broadband, and high-speed operation all come to fruition \cite{Marquez2020Graphene}. Generally speaking, this application of waveguides allows for the miniaturizing of photonic circuits while maintaining performance metrics required by AI hardware, pushing for significant improvement in the given domain \cite{wu2020two}.

\textbf{Nonlinear Optics} The strong nonlinear response of TMDCs also opens the door to advanced functions such as frequency conversion and all-optical signal processing. These capabilities are essential for implementing nonlinear optical functions directly on-chip as well as achieving on-chip quantum computing \cite{Pelgrin2023Hybrid}. 

\begin{table}[!htbp]
\centering
\caption{Second and third-order nonlinear optical parameters for common 2D materials and primary materials of CMOS-compatible platform used in Si and Si-hybrid integration schemes at technically significant telecom wavelengths. Characterizes nonlinear response of various hybrid waveguides demonstrating the promise in performance for 2D materials in current AI context. \cite{Pelgrin2023Hybrid}}
\label{tab:nonlinoptictbl}
\resizebox{\textwidth}{!}{
\begin{tabular}{lcccccc}
\toprule
Material & \begin{tabular}{c}
$n_2/n_{2\mathrm{Si}}$ \\
($n_{2\mathrm{Si}} = 7.2 \times 10^{-18}\ \mathrm{m}^2 \cdot \mathrm{W}^{-1}$)
\end{tabular} & \begin{tabular}{c}
Refractive index \\
(at 1550 nm)
\end{tabular} & TPA (at 1550 nm) [$\mathrm{m} \cdot \mathrm{W}^{-1}$] & 
\begin{tabular}{c}
$\chi^{(3)}$ [$\mathrm{m}^2 \cdot \mathrm{V}^{-2}$] \\
@ 1550 nm
\end{tabular} & 
\begin{tabular}{c}
$\chi^{(2)}$ [$\mathrm{m} \cdot \mathrm{V}^{-1}$] \\
(emission wavelength)
\end{tabular} & Band gap (eV) \\
\midrule
\rule{0pt}{4ex}Si & 1 & 3.5 & $0.5 \times 10^{-11}$ & $1.6 \times 10^{-21}$ & -- & 1.12 \\
\rule{0pt}{4ex}Si$_3$N$_4$ & 0.034 & 1.9--2.0 & None & $2.3 \times 10^{-22}$ & -- & -- \\
\rule{0pt}{4ex}SiO$_2$ & 0.003 & 1.44 & None & -- & -- & -- \\
\rule{0pt}{4ex}WS$_2$ & 291.6 & -- & $1.58 \times 10^{-9}$ & $2.4 \times 10^{-19}$ & $68 \times 10^{-11}$ (600 nm) & 2.1 \\
\rule{0pt}{4ex}MoS$_2$ & 
\begin{tabular}{c}
37.5 \\
15.28
\end{tabular} & -- & -- & $3.6 \times 10^{-19}$ & $2.9 \times 10^{-11}$ (780 nm) & 2 \\
\rule{0pt}{4ex}WSe$_2$ & -- & -- & -- & $1.0 \times 10^{-19}$ & $0.4 \times 10^{-11}$ (730 nm) & 1.75 \\
\rule{0pt}{4ex}MoSe$_2$ & -- & -- & -- & $2.2 \times 10^{-19}$ & $5 \times 10^{-11}$ (640--700 nm) & 1.7 \\
\rule{0pt}{4ex}BP & -- & -- & -- & $1.6 \times 10^{-19}$ & -- & 0.3--2 \\
\rule{0pt}{4ex}Graphene & 
\begin{tabular}{c}
2.08 \\
5.69 \\
10.69 \\
2.78
\end{tabular} & -- & -- & $1.0 \times 10^{-19}$ & -- & Zero-gap \\
\rule{0pt}{4ex}GO & 1600--3750 & -- & -- & -- & -- & -- \\
\bottomrule
\end{tabular}}
\end{table}

Graphene-based devices also highlights a promise for brain-inspired architectures like photonic neural networks - a recent study proposed a graphene-based synapse model embedded in microring resonators to enable large-scale neural networks using multi-wavelength techniques \cite{Marquez2020Graphene}, an approach that could significantly accelerate training processes for large language models.

\subsection{Case Study: AI Hardware Using Photonic Chips}

Photonic chips integrated with 2D materials are particularly promising for AI hardware due to their ability to perform computations faster than what current technology can achieve, even nearing the speed of light. For instance:

Researchers at MIT demonstrated a fully integrated photonic processor capable of executing deep neural network computations optically \cite{bandyopadhyay2024single}. By incorporating nonlinear optical function units (NOFUs), this chip achieves ultra-low latency while consuming minimal power, accomplishing the key computations for a machine-learning classification task in less than half a nanosecond while achieving greater than 92\% accuracy (matching the performance of current technology). This chip was also fabricated using commercial processes, allowing for reasonable scaling of this new technology \cite{bandyopadhyay2024single}.

\begin{figure}[h]
\centering
\includegraphics[width=1.\textwidth]{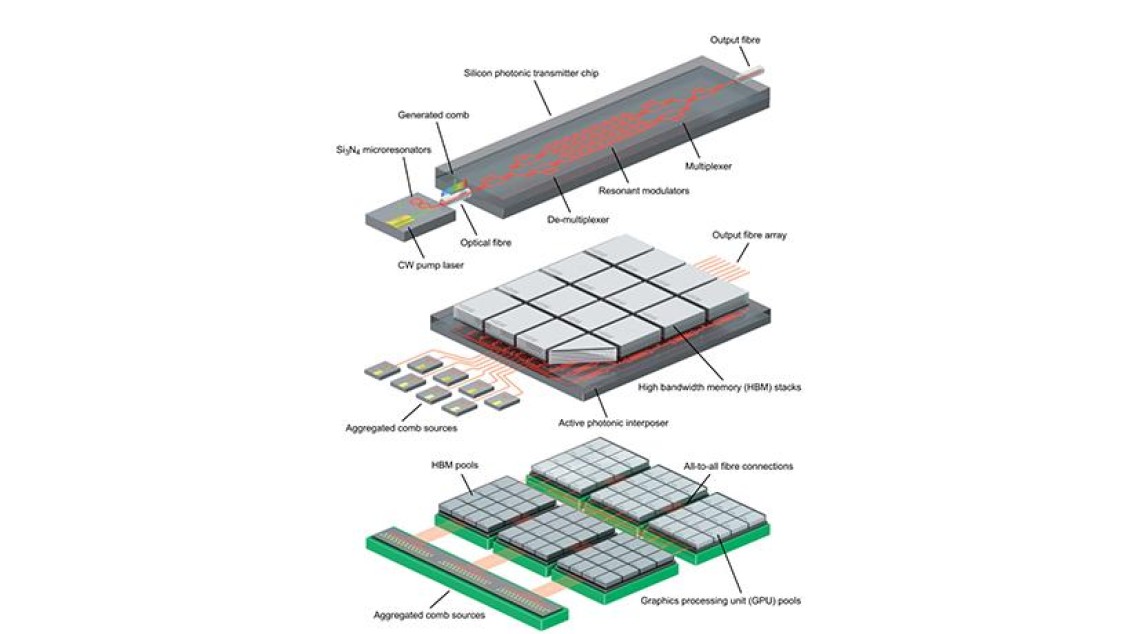}
\caption{Artistic depiction of the hierarchal structure based on Kerr frequency comb-driven silicon photonic links. 
\cite{rizzo2023massively}
}\label{fig:columbiaphotonics}
\index{figures}
\end{figure}

Columbia University developed an energy-efficient method for transferring larger quantities of data over fiber-optic cables, leveraging Kerr frequency combs on photonic chips, allowing researchers to send clear signals through separate and precise wavelengths of light \cite{rizzo2023massively}. This innovation improves bandwidth density while reducing energy consumption, both of which are critical factors in improving scaling of LLM training systems.

Black Semiconductor opened a new headquarters titled the FabONE facility focusing on developing graphene-based photonic connectivity solutions that unlock faster chip-to-chip interconnects. The technology here would enable advancements in high performance computing, AI, robotics, autonomous driving, and much more, especially ultrafast training processes for AI models \cite{fabonehq}.

These breakthroughs highlight the potential of 2D material-integrated photonic chips as they provide tangible results towards the revolutionization of AI infrastructure through addressing bottlenecks in speed, scalability, and energy efficiency.

\subsection{Challenges and Future Directions}

Despite their potential, as with all new technology, several challenges must be addressed to fully realize the benefits and tap into the future of 2D materials in integrated photonics:

\paragraph{Scalability} The fragility of ultrathin 2D materials poses challenges during large-scale manufacturing, and advances in transfer printing techniques and wafer-scale synthesis are needed to overcome this limitation to properly make this technology scalable \cite{Cao2021Recent}.
\paragraph{Material Stability} Some 2D materials, including both graphene and TMDCs, degrade under ambient conditions, and for this technology to catch on, there needs to be development of protective coatings, encapsulation techniques, or general preservation advancements for long-term reliability \cite{Vranic2025Biological}.
\paragraph{Integration Complexity} Achieving seamless integration with existing CMOS processes requires further optimization for varying techniques and interface engineering before this new technology can be properly integrated into the general world \cite{Pelgrin2023Hybrid}.

Future research should focus on addressing these challenges while continuining to explore new material systems that complement graphene and TMDCs. With both of these paths combined, the development of hybrid platforms combining electronic, photonic, and 2D material-based components seem promising in paving the way for transformative advancements in AI hardware and technology for the near future.

\section{Spintronics for Photonic Neuromorphic Computing Chips}
\subsection*{Introduction}
Nano-photonics, as an emerging interdisciplinary subject, integrates the principles of nanotechnology and photonics, aiming to explore and harness the manipulation of light wave by nanoscale structures. In the landscape of photonics, active devices and passive devices are crucial and have broad application prospects. Neuromorphic systems aim to mimic the computational and cognitive capabilities of the brain by leveraging the principles of neural networks. This section systematically investigates the synergistic integration of spintronic devices with nano-photonic architectures for neuromorphic computing.

\subsection{Background and Challenges of Neuromorphic Computing}

The pursuit of neuromorphic computing stems from fundamental limitations in conventional von Neumann architectures. Traditional computing systems suffer from the "von Neumann bottleneck" \cite{4}, where physical separation between processing and memory units leads to excessive energy consumption and latency during data transfer. This bottleneck is exacerbated by the growing performance gap between processors and memory, known as the "memory wall" \cite{5}. Modern computers require megawatts of power to simulate basic brain functions \cite{6}, while biological brains achieve remarkable cognitive capabilities with merely 20 W \cite{7}. Simultaneously, the semiconductor industry faces existential challenges as transistor miniaturization approaches physical limits and Moore's law stagnates \cite{8-10}. These dual crises in architecture and transistor scaling have driven intense interest in brain-inspired computing paradigms.

Neuromorphic computing addresses these challenges through three key innovations: 1) co-location of computation and memory, 2) analog information encoding, and 3) massively parallel connectivity \cite{11-16}. While theoretical frameworks for neural networks date back to McCulloch and Pitts' binary neuron model (1943) \cite{17} and subsequent developments in deep learning \cite{21,22}, practical implementations face significant hardware constraints. CMOS-based implementations using transistor arrays \cite{24} lack essential neurobiological features like nonlinear dynamics, long-term plasticity, and stochasticity \cite{16}. The emergence of nonvolatile memory technologies—particularly memristors \cite{25,26}—has enabled more biologically plausible implementations, but material limitations persist. Resistive RAM (RRAM) \cite{27-32}, phase-change materials \cite{33-37}, and ferroelectric devices \cite{38-42} face tradeoffs between endurance, speed, and controllability that constrain large-scale deployment.

Three generations of neural networks highlight evolving hardware requirements: 1) First-generation perceptrons \cite{20} with threshold operations, 2) second-generation deep neural networks (DNNs) \cite{21} requiring continuous nonlinear activation functions, and 3) third-generation spiking neural networks (SNNs) \cite{23} demanding precise temporal coding and event-driven processing. While DNNs dominate current AI applications, SNNs offer superior biofidelity and energy efficiency through sparse, spike-based communication \cite{13,23}. However, implementing SNNs in hardware remains particularly challenging, as it requires devices that inherently emulate biological neurons' leaky integrate-and-fire (LIF) dynamics \cite{51,52} and synapses' spike-timing-dependent plasticity (STDP) \cite{58}. Current solutions using CMOS circuits \cite{24} or emerging memristors \cite{25-42} either lack essential neuromorphic characteristics or suffer from limited endurance and stochastic control. This hardware-algorithm gap fundamentally restricts neuromorphic computing's potential to achieve brain-like efficiency and adaptability.

\subsection{Core Advantages and Key Spintronic Technologies for Neuromorphic Computing}

Spintronic devices exhibit unique advantages that position them as leading candidates for neuromorphic computing hardware. Their intrinsic nonvolatility, ultrafast dynamics (\SI{>1}{\giga\hertz}), and near-unlimited endurance (\num{e15} cycles) enable energy-efficient and biologically plausible neural network implementations \cite{Grollier2020}. Crucially, spintronic technologies leverage magnetic and spin-based phenomena to natively emulate neuro-synaptic functionalities while maintaining compatibility with conventional CMOS manufacturing processes \cite{Chen2021}. Three core advantages drive their prominence: (1) Stochasticity in magnetization switching and spin precession mirrors probabilistic neural firing mechanisms, enabling event-driven spiking neural networks (SNNs) with sparse coding efficiency \cite{Camsari2019}; (2) Multistate magnetization dynamics (e.g., domain wall motion, skyrmion nucleation) provide analog memristive behavior essential for synaptic weight modulation \cite{Locatelli2014}; and (3) Nonvolatile state retention eliminates static power consumption during idle periods \cite{Sengupta2017}. These attributes address critical von Neumann bottleneck limitations while surpassing competing memristive technologies in speed and reliability \cite{LeImini2020}.

The magnetic tunnel junction (MTJ) constitutes the foundational spintronic building block, demonstrating versatile neuromorphic functionality through two operational regimes. In superparamagnetic mode, stochastic switching between parallel and antiparallel states generates Poisson-distributed spikes for probabilistic computing \cite{Mizrahi2016}, achieving \SI{604}{\%} tunneling magnetoresistance (TMR) ratios in CoFeB/MgO structures \cite{Ikeda2008}. When configured as spin-torque nano-oscillators (STNOs), MTJs produce GHz-range voltage oscillations that synchronize with external stimuli, enabling coupled oscillator networks for pattern recognition \cite{Romera2018}. Spin-orbit torque (SOT) devices extend these capabilities through field-free magnetization switching in heavy metal/ferromagnet bilayers. SOT-driven spin Hall nano-oscillators (SHNOs) achieve mutual synchronization in 2D arrays \cite{Awad2020}, while three-terminal MTJs separate read/write paths for enhanced synaptic precision \cite{Fukami2016}. Domain wall motion in magnetic nanowires provides continuous resistance modulation ideal for analog synapses, demonstrating \SI{32}{\milli\electronvolt} energy per synaptic update \cite{Locatelli2014}.

Emerging topological spin textures like magnetic skyrmions offer particle-like dynamics for bio-inspired computing paradigms. Skyrmion nucleation and annihilation in chiral magnets (\SI{<100}{\nano\meter} diameter) emulate neurotransmitter release probabilities with \SI{10}{\micro\ampere} current thresholds \cite{Pinna2018}. Antiferromagnetic (AFM) spintronics introduces terahertz-range dynamics and stray-field immunity, enabling dense crossbar arrays through compensated magnetic moments \cite{Chen2019}. AFM-based synapses exhibit \SI{100}{\pico\second} switching speeds and thermal stability exceeding \SI{200}{\celsius} \cite{Wadley2016}. Integration of these technologies enables all-spin neural networks combining STNO-based neurons \cite{Romera2018}, domain wall memristive synapses \cite{Sengupta2017}, and skyrmionic probabilistic interconnects \cite{Pinna2018} – a hardware ecosystem addressing the memory-processor dichotomy through physics-level co-design.

\subsection{Exploration of System-Level Applications of Spintronics Technology}

Spintronic neuromorphic systems demonstrate transformative potential across cognitive computing paradigms through physics-enabled architectural innovations. A pioneering implementation employs four synchronized spin-torque nano-oscillators (STNOs) in coupled microwave emission states to achieve 96\% accuracy in real-time vowel recognition, outperforming equivalent deep learning networks by 17\% while consuming 3~mW per classification \cite{46}. This event-driven architecture leverages the intrinsic frequency multiplexing of 2.4~GHz STNO arrays to map temporal speech patterns directly onto oscillator synchronization states, eliminating analog-to-digital conversion overhead \cite{69}. For large-scale implementations, spin Hall nano-oscillator (SHNO) crossbars with 32$\times$32 elements demonstrate mutual phase locking across 100~μm distances through propagating spin waves, enabling pattern completion tasks through collective dynamics rather than discrete synaptic weights \cite{47}. 

Magnetic skyrmion fabrics introduce probabilistic computing capabilities through topologically protected particle interactions. Networks of 50--100~nm skyrmions in chiral magnets implement Bayesian inference engines by encoding probability distributions in nucleation density, achieving 92\% accuracy in weather prediction models through in-memory sampling of 10\textsuperscript{5} stochastic states \cite{44}. This approach reduces Monte Carlo simulation energy by 10\texttimes compared to GPU implementations through analog current-controlled reshuffling \cite{166}. Antiferromagnetic (AFM) spintronics enables ultra-dense architectures through stray-field immunity and 1~THz dynamics, with experimental 4~fJ per synaptic update in IrMn-based crossbars maintaining 0.1\% weight drift over 10\textsuperscript{12} cycles \cite{143,145}. 

Reservoir computing implementations exploit nonlinear magnetization dynamics for temporal signal processing. A single vortex-STNO achieves 400-neuron equivalence through time-multiplexed precession states, solving Mackey-Glass chaotic time series prediction with 0.012 normalized mean square error \cite{69}. Skyrmion-based reservoirs leverage emergent interactions in disordered magnetic textures to process 10~MHz EEG signals with 20~μW power consumption, demonstrating real-time epileptic seizure detection through bifurcation detection in spin texture dynamics \cite{167}. Looking toward large-scale deployment, all-spin neural networks combining STNO neurons \cite{46}, domain wall synapses \cite{80}, and AFM interconnects \cite{143} promise >100~TOPS cognitive performance at <10~mW system power through physics-level co-design of neuro-synaptic functionalities.

\begin{figure}[h]
\centering
\includegraphics[width=1.\textwidth]{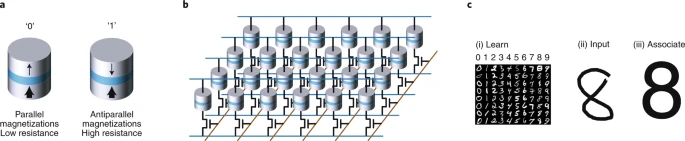}
\caption{Magnetic tunnel junctions for memory applications. a, A magnetic tunnel junction consists of two ferromagnetic layers (grey) separated by an
insulating layer (blue) with the magnetization of one layer fixed and that of the other either parallel (low resistance) or antiparallel (high resistance) to
it. The labels ‘1’ and ‘0’ indicate the configurations that represent each value. b, Cross-bar array of magnetic tunnel junctions for high-density storage
(magnetic random-access memory). The resistance of a particular tunnel junction is measured by activating the appropriate word line (red) allowing
conduction between the bottom bit line and the top sense line (both blue). The alignment of the magnetization can be switched by passing sufficient
currents through the device. c, Associative memory: (i) handwritten digits from the MNIST dataset used for training the associative memory; (ii) sample
test input after training; (iii) output of trained network from the test input showing successful association. 
\cite{}
}\label{fig:sixuan3}
\index{figures}
\end{figure}

\begin{figure}[H]
\centering
\includegraphics[width=1.\textwidth]{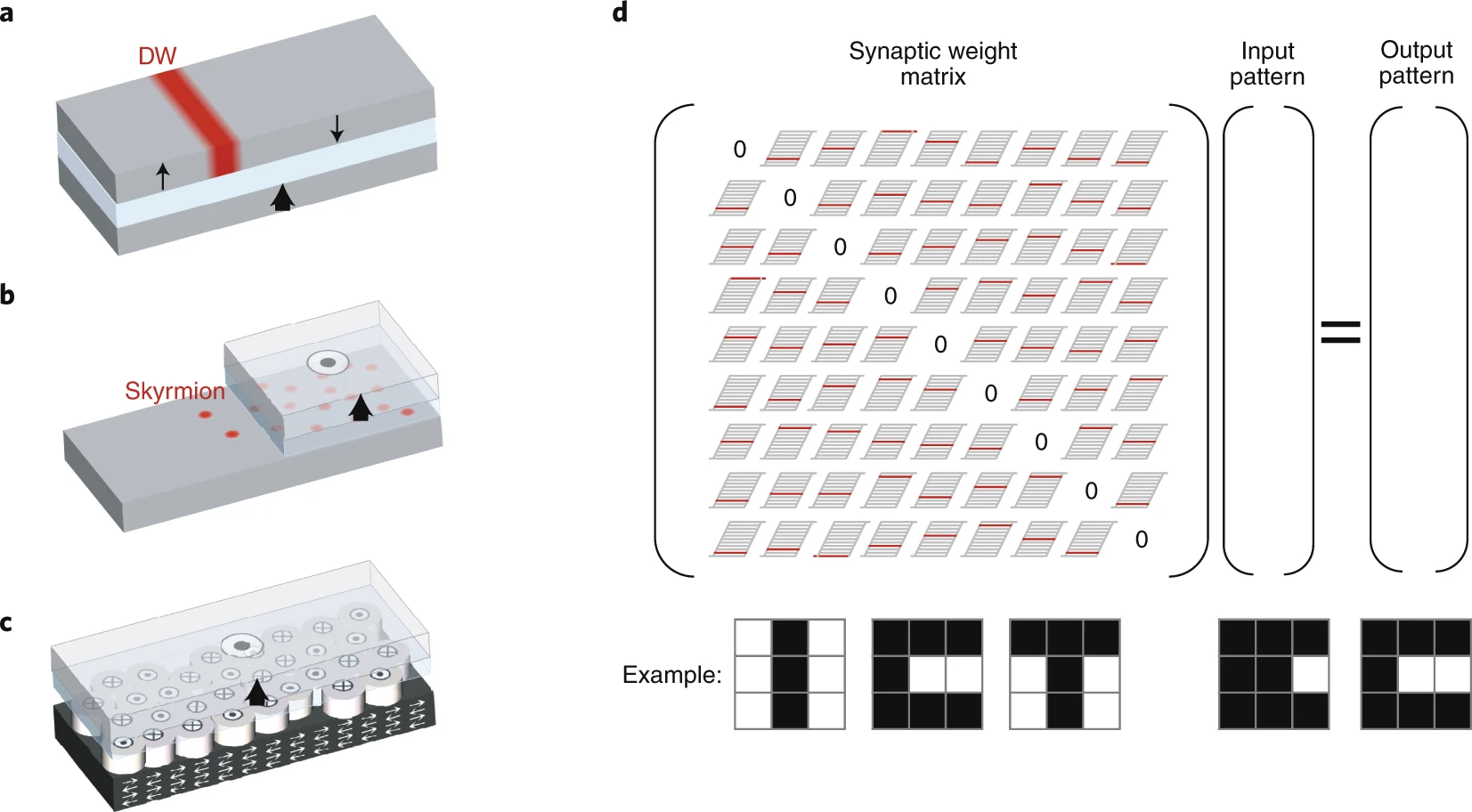}
\caption{| Spintronic-based memristors. a, Domain-wall memristor. The resistance of the magnetic tunnel junction depends on the location of the domain
wall changing the relative area of the high-resistance antiparallel configuration and the low-resistance parallel configuration. b, Skyrmion-based memristor.
The resistance of the device depends on the number of skyrmions under the fixed layer. c, Fine-magnetic-domain tunnelling memristor. In a tunnel junction
coupled to a polycrystalline antiferromagnet, the variation of switching properties from domain to domain allows the domains to reverse independently
and under different conditions. The resistance of the device then depends on the fraction of domains with magnetizations aligned with the uniformly
magnetized fixed layer. d, Spintronic associative memory. The value of each off-diagonal matrix element is stored in the configuration of the memristor
schematically illustrated by the different levels in the matrix. These levels are trained so that when the matrix multiplies an input, the result is the closest
element of the training set. The multiplication is carried out by applying voltages that corresponding to the input and measuring the output current through
the appropriate memristors. The first three example images at the bottom of d are the three images that the network has been trained to recognize.
The fourth is a ‘noisy’ version of one of these images and the fifth is the reconstructed correct image.
\cite{}
}\label{fig:sixuan3}
\index{figures}
\end{figure}

\section{Principles and Mechanisms of Transformer Neural Networks and LLMs}

\begin{figure}[hbtp]
\centering
\begin{subfigure}[b]{0.4\textwidth}
    \includegraphics[width=\textwidth]{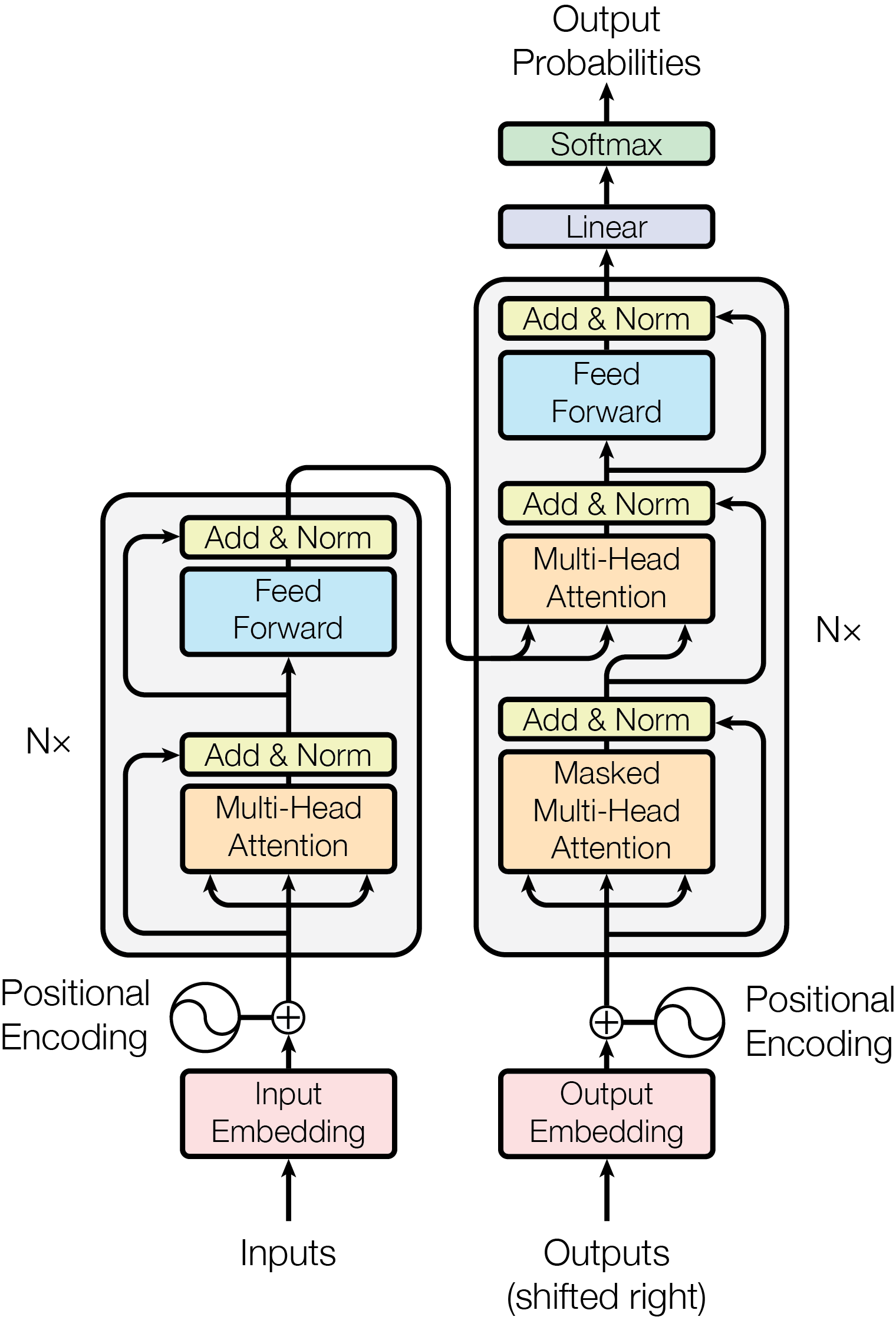}
    \caption{Schematic of the scaled dot-product attention mechanism. Queries (Q), keys (K), and values (V) are each transformed by learned linear projections. The dot products of Q and K are scaled, passed through a softmax, and multiplied by V to produce attention outputs. Multiple parallel heads capture diverse token relations, and their outputs are concatenated and linearly projected again.}
\end{subfigure}
\hfill
\begin{subfigure}[b]{0.4\textwidth}
    \includegraphics[width=\textwidth]{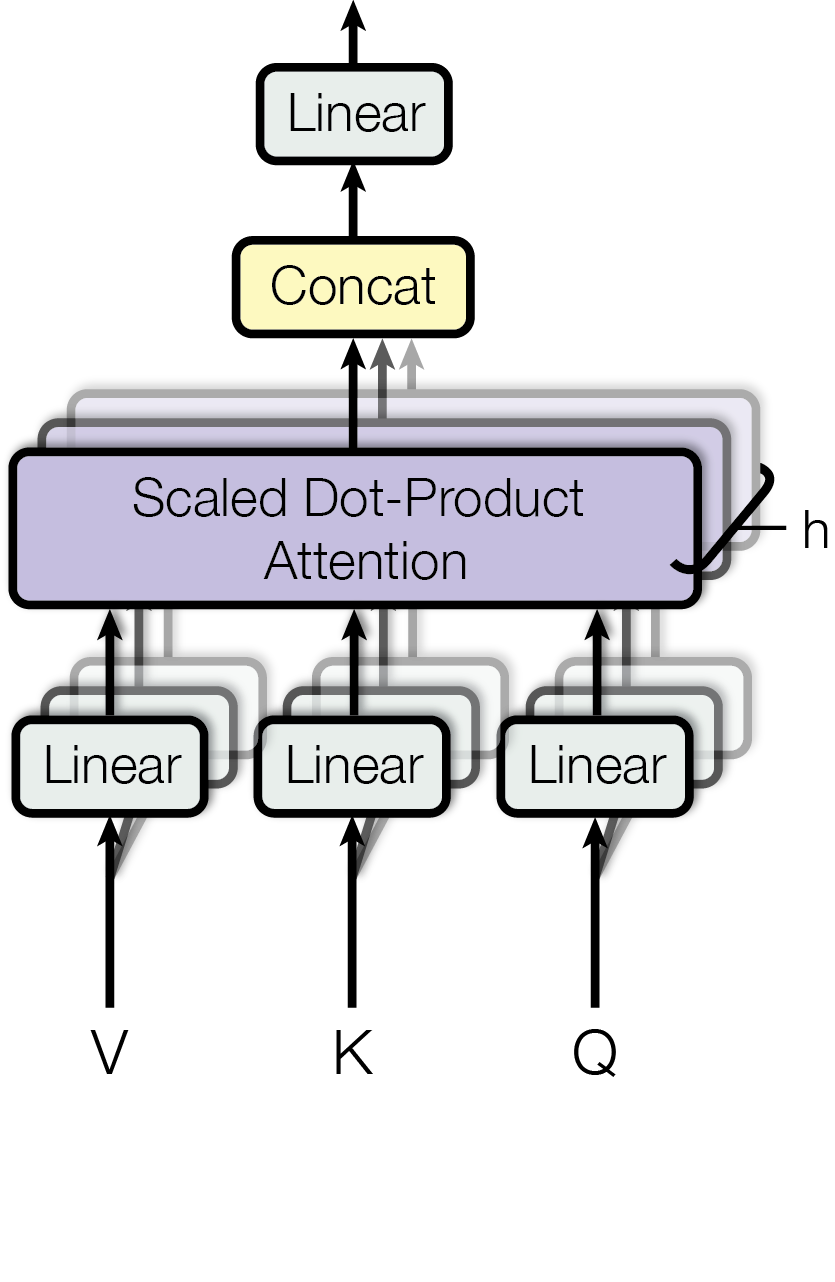} 
    \caption{High-level view of the Transformer architecture \cite{vaswani2017attention}, illustrating the encoder (left) and decoder (right). The encoder stack consists of multi-head self-attention and feed-forward layers, while the decoder includes masked multi-head attention for autoregressive generation plus encoder-decoder attention to focus on encoder outputs.}
\end{subfigure}
\caption{Transformer neural networks used in modern LLMs. (a) The scaled dot-product attention mechanism that underpins multi-head attention in Transformers. (b) The original Transformer design featuring an encoder-decoder structure for sequence-to-sequence tasks. Reproduced with permission.\cite{vaswani2017attention}}
\label{fig:transformer_overview}
\index{figures}
\end{figure}

\subsection{Transformer achitecture}
Existing LLMs are all based on a DNN proposed by Vaswani et al. \cite{vaswani2017attention} who introduced sequence modeling by relying on an attention mechanism instead of recurrence or convolution, which is now widely known as the \textit{Transformer} architecture \cite{vaswani2017attention}. In the original Transformer design for machine translation \cite{vaswani2017attention}, an encoder-decoder structure was employed. The encoder stack processes the input through self-attention layers—which allow each token to attend to others in the sequence—followed by position-wise feed-forward networks. The decoder stack then generates output tokens using a self-attention mechanism combined with encoder-decoder attention to focus on the encoder’s output \cite{vaswani2017attention}. This design enables the model to handle sequences without maintaining an RNN-style hidden state, thereby improving parallelization during both training and inference \cite{vaswani2017attention}. With this architecture, the Transformer achieved superior translation quality while requiring significantly less time to train compared to prior recurrent or convolutional models \cite{vaswani2017attention}.

The key innovation behind the Transformer is \emph{self-attention}. This mechanism helps build contextualized representations by allowing each position in the sequence to selectively attend to other positions. Another essential idea is \emph{multi-head attention}, which is used to capture different aspects of token relations \cite{vaswani2017attention}. Each attention head learns to focus on different patterns, enabling the model to integrate diverse information about word relationships by combining the outputs from multiple heads \cite{vaswani2017attention}. This attention mechanism gives the Transformer the ability to process long sequences effectively; since any token can influence any other through weighted attention, it addresses long-range dependencies more robustly than the fixed-step interactions of RNNs. Furthermore, Transformers are highly scalable to long sequences because attention across all positions can be computed in parallel. In contrast, recurrent neural networks must process tokens sequentially.

Another important concept in Transformers is the incorporation of \emph{positional encodings} to inject information about token positions into the model. The original approach used fixed sinusoidal position embeddings added to token embeddings \cite{vaswani2017attention}. These positional signals help the model understand the ordering of words (e.g., distinguishing “Alice answered Bob” from “Bob answered Alice”). After embedding the inputs and adding positional encodings, each Transformer layer applies layer normalization and residual skip connections around its sub-layers. The residual connections mitigate vanishing gradient issues by adding the layer’s input to its output, while layer normalization ensures that activations remain well-conditioned.

Together, the Transformer architecture—comprising multi-head self-attention, feed-forward networks, residual/normalization layers, and positional encodings—provides a highly parallelizable and effective approach to modeling sequences. It quickly became the dominant architecture in natural language processing, enabling the training of much larger models than was feasible with RNNs or CNNs, thanks to its ability to capture long-range context and process entire sequences in parallel \cite{language_models_unsupervised_multitask}.

\subsection{Chain-of-Thought Prompting and Reasoning}
When thinking step-by-step, large language models exhibit emergent reasoning capabilities. \emph{Chain-of-Thought (CoT) prompting} is a technique in which the model is explicitly guided to generate a sequence of intermediate reasoning steps before producing a final answer. In this approach, the model is instructed to provide intermediate reasoning steps that effectively decompose a complex problem into simpler sub-problems, thereby creating a solution path. CoT has been shown to dramatically improve performance on arithmetic, commonsense, and symbolic reasoning tasks \cite{cot2022reasoning}. For example, Wei et al. \cite{cot2022reasoning} prompted a 540-billion-parameter model (PaLM) with a few step-by-step exemplars and achieved state-of-the-art accuracy on the GSM8K math word problem set, even surpassing a fine-tuned 175B GPT-3 model. Figure 17 below vividly illustrates how a standard prompt yields an incorrect one-line answer, whereas a CoT-prompted model produces a multi-line explanation that arrives at the correct answer.

\begin{figure}[hbtp]
\centering
\includegraphics[width=\textwidth]{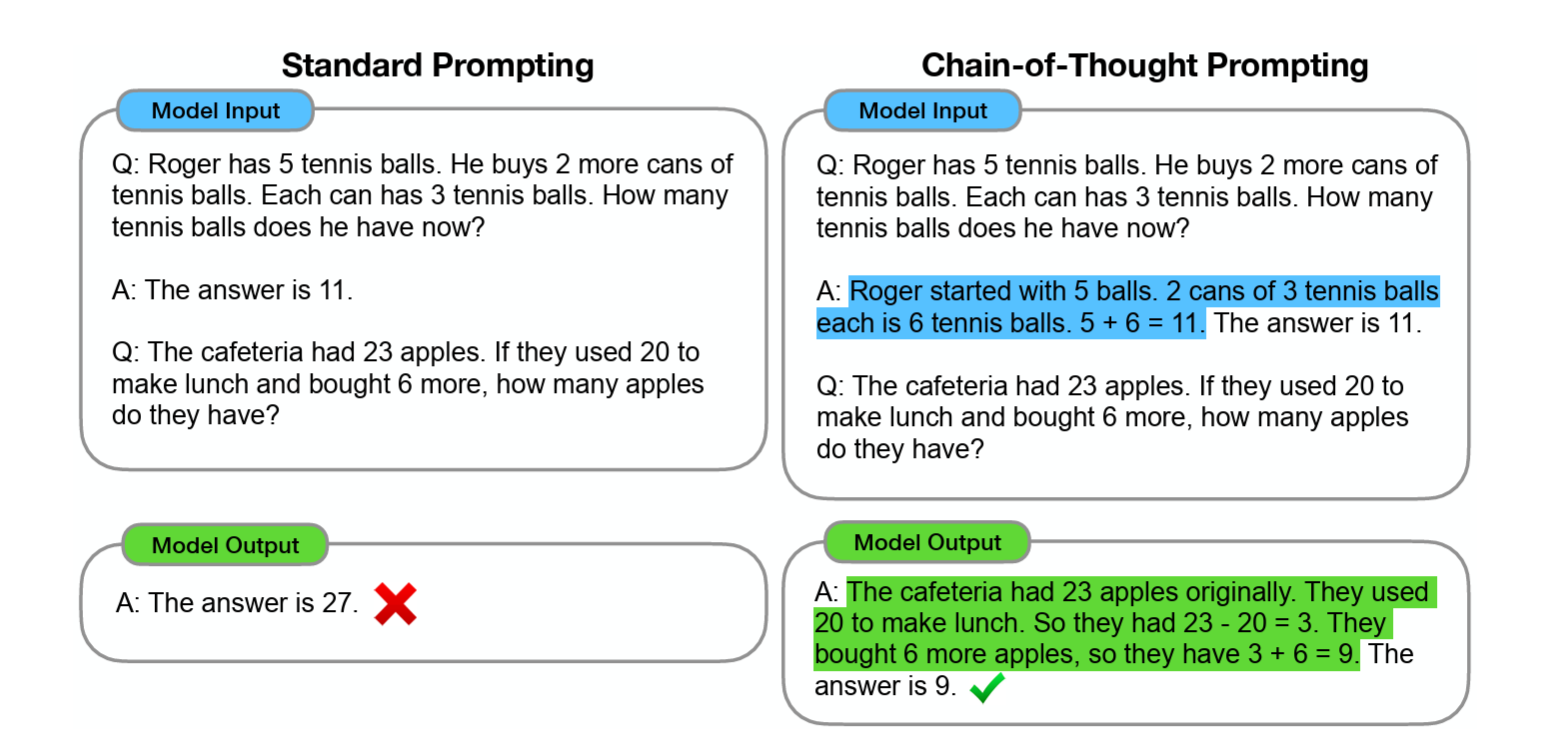}
\centering
\caption{Chain-of-thought prompting enables large language models to tackle complex arithmetic, commonsense, and symbolic reasoning tasks. Chain-of-thought reasoning processes are highlighted.\cite{shinn2023reflexion}.}
\label{fig:COT}
\index{figures}
\end{figure}

The integration of CoT can also be achieved via training. Few-shot CoT prompting uses a small set of manually crafted Q\&A examples with detailed rationales to train the model \cite{cot2022reasoning}. More recently, models are being fine-tuned on datasets comprising questions paired with their step-by-step solutions, enabling the model to generate a chain of thought even in zero-shot settings. For instance, the STaR method (Self-Taught Reasoner) allows the model to generate its own reasoning on unlabeled problems and verify its answers. Another technique, known as self-consistency, involves generating multiple distinct chains-of-thought and then selecting the most common answer among them to reduce the likelihood of an incorrect reasoning path. Recent developments suggest that chain-of-thought has even become an architectural consideration in large language models, with Google’s PaLM 2 and OpenAI’s GPT-4 believed to leverage internal CoT-style prompting during training or via RLHF. Appending phrases such as “Let’s think step by step” to a prompt (a form of zero-shot CoT) often triggers the production of a reasoning trace, thereby improving accuracy on complex tasks \cite{llm_zero_shot}.

\subsection{Self-Reflection and Self-Critique in LLMs}
An innovative mechanism in advanced LLMs is self-reflection, which enables models to analyze and refine their outputs. Typically implemented via reflective prompting and feedback loops, the model first generates an initial solution with a chain-of-thought, and then, through subsequent prompts, critiques its own reasoning to identify and correct errors. Studies have demonstrated significant improvements in problem-solving performance when such a self-reflection mechanism is employed \cite{selfreflection2023}. This can be done iteratively – each reflection fed advice (or an inner prompt) back into the model for another try \cite{qu2024recursive}. Such a feedback loop serves as a “debugging tool” for the model.

One implementation, known as \emph{Reflexion}, treats the model’s output as a source of feedback with dynamic memory to improve reasoning scores. In Reflexion, after each action, the model receives a binary reward (success/failure) and can record a brief reflection about what went wrong or could be improved. Then these reflections are appended to an episodic memory and used to guide the next trial. For example, a heuristic function works if the model encounters an error or a dead-end (hallucination or inefficient trajectory), then the model may reset and restart the task with the benefit of hindsightThis trial-and-error loop, similar to an RL fine-tuning process, led to improved performance on benchmark tasks by effectively letting the model learn from past experience in natural language. Shinn et al. \cite{shinn2023reflexion} showed that a GPT-4-based agent using Reflexion achieved 91\% accuracy on the HumanEval coding benchmark.

\subsection{Reinforcement Learning from Human Feedback (RLHF) for Alignment}
Reinforcement Learning from Human Feedback (RLHF) aligns large language models with human intents and values. This technique uses a reward signal from human preference judgments, thus resulting in fine-tuning the model\cite{rhlf2022}. The architecture for RLHF involves a feedback pipeline with a few components: (1) a supervised fine-tuning (SFT) stage to teach the model basic desired behavior (2) a reward model that scores outputs based on human preferences (3) a policy optimization stage (often using reinforcement learning algorithms like Proximal Policy Optimization, PPO) to train the model to maximize the learned reward.

In the case of OpenAI’s GPT-4 the process worked roughly as follows:
\begin{enumerate}
  \item \textbf{Supervised Alignment Tuning:} The base model is fine-tuned on a set of demonstrations of good behavior. Human annotators craft datasets with example prompts and correct responses. The base model learns to produce human reaction outputs (e.g. polite, accurate answers following the user’s request).
  \item \textbf{Reward Model Training:} Human evaluators rank multiple model outputs for a variety of prompts, creating a dataset used to train a reward model $R(\text{prompt, answer})$ that outputs a scalar score reflecting human preference.
  \item \textbf{Policy Optimization:} The fine-tuned model from Step 1 is further fine-tuned using an RL algorithm to maximize the reward model’s score. PPO (Proximal Policy Optimization) is often used in this stage, where the model’s parameters are adjusted to increase the probability of high-reward outputs after the model generates outputs for a given prompt with scores from the reward model. Kullback–Leibler (KL) divergence penalty is often implemented in the reward objective to prevent degeneration or excessive safe answers. Through many iterations of this process on a wide variety of prompts, the model learns to output responses that align with human preferences.
\end{enumerate}
The architecture of RLHF thus includes the optimization loop connecting between the original model (policy) and a reward model. This can be compared to human human-provided training signal rather than a fixed labeled dataset. Other than the classic RLHF loop, Direct Preference Optimization (DPO), was introduced by Rafailov et al. (2023), is another emerging alternative. DPO reformulates the preference alignment task as a direct supervised objective, removing the need for a separate reward network and RL optimization \cite{dpo_reward}. DPO treats the large language model itself as “secretly a reward model” and fine-tunes it on the comparison data in closed form. For example, DeepSeek-LLM utilized DPO in the RLHF phase), which surpass GPT-3.5 on various open-ended tasks. RLHF via PPO requires carefully balancing the reward maximization to avoid over-optimization\cite{deepseek}.

\begin{figure}[hbtp]
\includegraphics[width=0.55\textwidth]{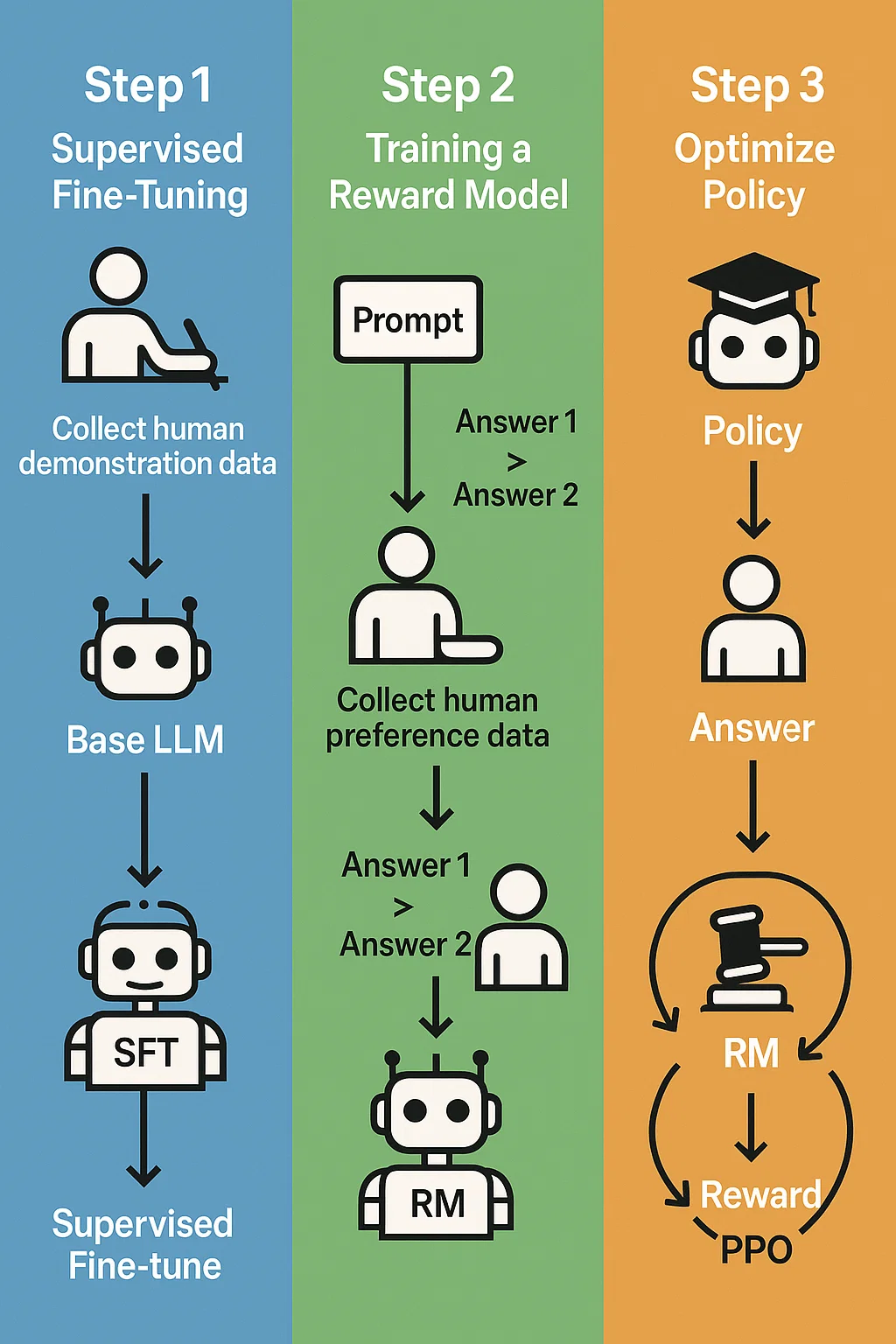}
\centering
\caption{The core of RLHF is training a separate AI reward model based on human feedback, and then using this model as a reward function to optimize policy through RL. Given a set of multiple responses from the model answering the same prompt, humans can indicate their preference regarding the quality of each response. You use these response-rating preferences to build the reward model that automatically estimates how high a human would score any given prompt response. The language model then uses the reward model to automatically refine its policy before responding to prompts. Using the reward model, the language model internally evaluates a series of responses and then chooses the response that is most likely to result in the greatest reward. This means that it meets human preferences in a more optimized manner. }
\label{fig:self}
\index{figures}
\end{figure}

\subsection{Tool Use and Function Calling in LLMs}
Another frontier in LLM development is enabling models to use external tools or APIs. For example, LLM can call a calculator for math, a search engine for up-to-date info, or even customized functions. This approach gives LLM the ability to act as an agent that can query other systems and incorporate the results into its reasoning. Two notable examples are Toolformer (from Meta AI) and function calling in Qwen/GPT-4.

Toolformer (Schick et al., 2023) demonstrated that language models can be trained to decide when to call an API, which API to call, with what arguments, and how to incorporate the result into the text \cite{toolformer2023}. Toolformer was trained in a self-supervised way. Starting from a few manual examples of tool use, Toolformer generated synthetic data by inserting API calls into responses where they reduced the model’s perplexity. This gives a Transformer model ability to interact with other tools with normal text generation. For instance, it can output a token sequence indicating API calling(e.g., \texttt{[CALL: WikiSearch('Brown Act')]}). After receiving the result from the external tool, it can continue text generation includes the factual content from that result. Toolformer was shown to significantly improve zero-shot performance on tasks like arithmetic, question answering, and translation by leveraging tools. From the Transformer architecture perspective, Toolformer extends the decoder to produce special tool-use tokens. Thus, the model’s output space is augmented with actions. The system halts decoder actions when the model outputs a tool API token. After executing the API call, the result will be inserted into the model’s context of text generation.  The training procedure (illustrated in Figure 2 of the paper) involves sampling possible tool calls in a context, executing them, and filtering those that actually improve next-token prediction before adding them to the training data. Through this, the model learns where tool use is helpful and how to format the calls.\cite{toolformer2023}.

\begin{figure}[hbtp]
\centering
\includegraphics[width=\textwidth]{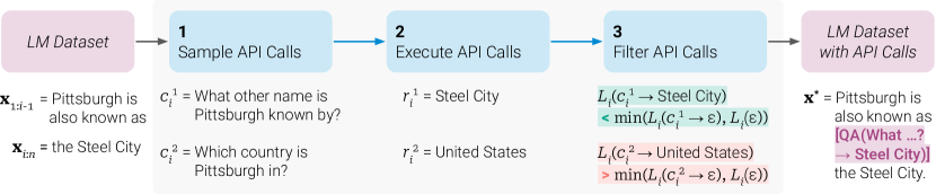}
\centering
\caption{Tool use demonstration. Illustrated is for a question answering tool: Given an input text \(x\), we first sample a position \(i\) and corresponding API call candidates \(c^{i}_1, c^{i}_2, \dots, c^{i}_k\). We then execute these API calls and filter out all calls which do not reduce the loss \(L_i\) over the next tokens. All remaining API calls are interleaved with the original text, resulting in a new text \(x'\).
\cite{toolformer2023}}
\label{fig:self}
\index{figures}
\end{figure}

\subsection{Memory Integration for Long-Term Context}
In the early stage, Transformer models were limited to input sequences of only a few thousand tokens due to the constraints of positional encoding schemes and the \(O(n^2)\) memory cost of self-attention. However, long-context reasoning in advanced large language models requires architectural innovations to handle long input sequences efficiently. Such innovations include enhanced positional encodings (e.g., ALiBi and Rotary embeddings) as well as efficient attention algorithms like FlashAttention.

\paragraph{Positional Encoding and Extrapolation:}  
Transformers require a mechanism to encode the position of each token since the model itself is order-agnostic. Standard methods---such as fixed sinusoidal embeddings---have a fixed context limit and often struggle to generalize beyond the training length. \emph{ALiBi} (Attention with Linear Biases) is a technique that enables extrapolation to longer sequences without retraining by dispensing with absolute positional embeddings. Instead, ALiBi adds a fixed penalty proportional to the distance between the query and key tokens, introducing a linear bias in the attention scores \cite{alibi2021}. The attention mechanism naturally down-weights far-distance tokens but never entirely ignores them. Press et al. \cite{alibi2021} showed that a model trained with ALiBi on 1K-token sequences can generalize to 2K or more tokens during inference, achieving performance comparable to a model trained on longer sequences. In essence, ALiBi allows long-context reasoning without a fixed positional index limit, as tokens beyond the training length simply receive larger bias values rather than entirely novel embedding vectors.

Another popular approach is \emph{Rotary Position Embedding (RoPE)} \cite{rope2021}. RoPE encodes positional information by rotating the query and key vectors within each attention head using a rotation matrix defined by sinusoidal frequencies. This rotation angle increases with the token’s position, enabling the inner product of rotated queries and keys to depend solely on their positional difference rather than on absolute positions. Consequently, the model is able to process sequences longer than those encountered during training. Empirically, models using RoPE have been scaled from 2K to 8K or even 16K-token contexts through interpolation methods. Jianlin \cite{rope2021} demonstrated that models incorporating RoPE achieved improved performance on long-sequence benchmarks compared to alternative approaches.

\paragraph{Efficient Attention Computation:}  
Even with enhanced positional encoding, the standard self-attention mechanism has a quadratic \(O(n^2)\) memory cost when processing large amounts of tokens. \emph{FlashAttention} \cite{flashattention2022} is an exact attention computation method optimized to minimize memory reads and writes, effectively making the computation I/O-bound rather than memory-bound. FlashAttention introduces a tiling strategy to store intermediate results in high-speed on-chip memory, dramatically reducing the need for expensive GPU memory access. Thus, while naive self-attention scales quadratically with sequence length, FlashAttention uses memory linear in the sequence length and achieves significant speedups.


Another recent technique is NTK (Neural Tangent Kernel)-aware interpolation, which “stretches” the rotary positional embeddings during inference. This enables an extension of the context length without retraining. Alibaba’s Qwen-7B/14B models, for example, utilize NTK-aware interpolation, log-scaled attention, and local window attention to achieve context lengths well beyond 8K tokens \cite{bai2023qwen}. The underlying principle of RoPE is that when frequencies are not modified, extending the trained context beyond boundaries results in angles that the model has never encountered. However, NTK interpolation ensures the rotation angles are adjusted within a range that the model is familiar with the new maximum length. This stretching technique has been empirically demonstrated in code-generation models (e.g., CodeLlama was extended from 16K to 100K with minimal performance degradation using a similar approach).

Advanced LLMs employ several mechanisms to enable long-context reasoning. These include positional encoding techniques that facilitate extrapolation—such as ALiBi’s distance-based linear bias and RoPE’s rotational encoding—and efficient attention algorithms like FlashAttention that overcome the quadratic memory bottleneck. In combination, these mechanisms have significantly increased the effective context length from roughly 1K to as high as 100K tokens, thereby expanding the range of applications from reading long contracts and logs to maintaining coherent context over vast document collections. Long-context architectures are thus crucial for bringing LLMs closer to human-like long-term coherence in conversation and writing.

\subsection{ChatGPT (OpenAI GPT-3.5/4 based)}
ChatGPT is one of the most well-known LLMs developed by OpenAI. It is optimized for dialogue via alignment training, making it a fine-tuned version of the GPT-3.5/GPT-4 family. ChatGPT’s underlying architecture is a decoder-only Transformer.  The model consists of multiple layers of self-attention and feed-forward networks, which enables it to capture long-range dependencies in text. It uses the transformer block as the GPT series (multi-head attention, layer normalization, residual connections). To maintain extended window, ChatGPT uses large context window, and the model uses positional encodings to handle sequential context. Notably, ChatGPT possesses the ability to follow a chain-of thought reasoning process when prompted. This enables it to effectively perform multi-step logic by internally generating reasoning steps before providing final answers \cite{cot2022reasoning}. This capability is attributed to the underlying architecture’s ability to maintain long contexts with the architecture of standard Transformer decoder \cite{gpt4technical}.

ChatGPT’s success is not attributed to novel architecture, but to innovations in training and prompting. A key innovation is utilization of RLHF for alignment, which was one of the first skill aligning a large language model with human feedback signals. Compared to the original GPT—which was a generic next-word predictor—ChatGPT was supervised with human feedback using instruction-based examples, and further refined through Reinforcement Learning from Human Feedback (RLHF) \cite{gpt4technical}. In the RLHF phase, the model was tuned with a reward model and policy optimization to prefer higher-ranked responses, resulting in more factual and polite outputs.  Another innovation is ChatGPT possesses the ability to utilize zero-shot and few-shot prompting, enabling it to adapt to tasks without the need for additional parameter updates, thereby inheriting GPT-3’s in-context learning capabilities. Wei et al. \cite{cot2022reasoning} points out CoT prompting can unlock enhanced reasoning performance on arithmetic, commonsense, and multi-hop reasoning problems. In CoT prompting, the model is guided to produce step-by-step intermediate reasoning before final answers to exhibit robust performance. Although this prompting framework is not an architectural change, it represents a technical refinement of reinforcement learning that enhances the reliability and controllability of the model’s outputs. Additionally, ChatGPT has iterative deployment and feedback, which is a feedback loop between the model and users. It collects interactions with real users and utilizes the data to further refine the model’s behavior. Coupled with RLHF and advanced prompting, this approach constitutes the primary technical advancements that transformed a generic LLM into a conversational agent \cite{gpt4technical}.

\subsection{LLaMA}
LLaMA is a family of open-source LLMs released by Meta in 2023, renowned for achieving high performance with smaller model sizes. The original LLaMA models were offered in sizes ranging from 7B to 65B parameters, trained on a mix of text from 20 languages. All versions share the same architectural template. The largest 65B model has the highest number of layers and hidden dimensions, while the smallest 7B model is significantly narrower. Meta found that the LLaMA-13B model, in particular, outperformed OpenAI’s GPT-3 (175B) on several benchmarks, highlighting the benefits of high-quality training data and extended training runs \cite{llama2}. The LLaMA models are built on the Transformer decoder architecture. In terms of architecture, LLaMA does not use standard transformer formula, and each LLaMA model is a decoder-only Transformer comprising a stack of self-attention layers and feed-forward layers. For instance, LLaMA employs rotary positional embeddings in its self-attention mechanism instead of absolute positional encodings, improving the ability to handle extended context lengths by rotating queries and keys in each attention head. The activation function in the feed-forward layers utilizes the SwiGLU variant (a gated linear unit activation that has been demonstrated to enhance training stability and performance, as employed in PaLM by Chowdhery et al. \cite{chowdhery2023palm}), providing a slight enhancement over the GELU activations. 

While LLaMA did not modify modern architectural components, it combined several technical refinements that contributed to its performance and efficiency. One technique is the meticulous curation of the training corpus. Low-quality or duplicate texts were rigorously filtered, resulting in a cleaner dataset that enables the model to concentrate on learning valuable patterns. LLaMA employed a SentencePiece tokenizer optimized for the training data, with a vocabulary of approximately 32k tokens \cite{llama_meta}. The tokenizer was designed to handle multiple languages and code languages since the corpus was multilingual. This attention to training data quality and tokenization improved the signal-to-noise ratio performance during training. 

Within the Transformer architecture, LLaMA incorporated several modifications from prior research. Two notable changes are Rotary Positional Embedding(RoPE) and SwiGLU activation function. Compared to original position embeddings, RoPE enables the model to have better generalization to longer sequences, while SwiGLU provides a modest performance improvement without incurring additional computational costs \cite{rope2021}. Additionally, LLaMA employs pre-normalization, which involves applying layer norm prior to attention and feed-forward sublayers. This technique has been demonstrated to enhance the stability of training deep transformers.
Overall, the LLaMA series demonstrates that with efficient training strategies, smaller-scale decoder-only Transformers can achieve competitive performance while maintaining openness.

\subsection{DeepSeek (Open-Source 67B Model)}
DeepSeek is an open-source LLM introduced in 2024, notable for its focus on bilingual capability and domain expertise. The DeepSeek (67B) model was trained from scratch on an extremely large dataset of approximately 2 trillion tokens in both English and Chinese \cite{deepseek}. DeepSeek base model made substantial modification on Transformer architecture to improve efficiency and scalability. 

DeepSeek introduces sparsely activated layers through a Mixture-of-Experts (MOE) \cite{bi2024deepseek} design to substitute feed-forward layers.  Instead of a single dense feed-forward network per Transformer block, there are multiple expert networks and a learned gating function selects which few experts handle each token’s transformation. This approach reduces the total number of parameters while maintaining computational efficiency \cite{bi2024deepseek}. This can reduce computational costs for inference or training on per-token basis to achieve the capacity of a super-large model.

Another architectural innovation in DeepSeek is Multi-Head Latent Attention (MLA) \cite{bi2024deepseek}. This technique was introduced to address the memory and speed bottlenecks of standard multi-head attention. MLA compresses the key and value representations in the attention mechanism by projecting them into a lower-dimensional latent space. Each attention head operates on a compressed representations, which is subsequently expanded as necessary. This reduces memory usage during training and inference on expanded sequences. In DeepSeek-V3, this method was further refined utilizing low-rank projections that adapt the compression based on the content.

DeepSeek’s architecture heavily relies on low-precision computation and custom optimization. The models were trained using 8-bit floating point (FP8) precision for most calculations, facilitated by software and hardware co-design. The team developed a scheme to maintain model quality when using FP8 for matrix multiplications. Additional, Deepseek introduced a custom parallelization strategy called DualPipe, which optimally overlaps communication and computations. These optimizations enabled DeepSeek to train models with fewer GPUs and weaker hardware. 

DeepSeek-R1 was built upon the DeepSeek-V3 architecture, which incorporates MoE and MLA with further prepared reinforcement learning training \cite{guo2025deepseek}. Architecturally, R1 does not introduce new layers but with additional conditioning, such as value-head outputs for reinforcement learning or policy networks. In summary, DeepSeek’s architecture can be characterized as a next-generation Transformer that prioritizes sparsity and computational efficiency. By integrating Multi-Head Attention (MoE) for parameter sparsity and Multi-Layer Perceptron (MLA) with low precision, the architecture achieves high capacity at a reduced computational cost.

\section{Principles and Algorithms of Spiking Neural Networks and Their Applications}

In recent years, spiking neural networks (SNN) have emerged as a powerful AI paradigm to efficiently realize LLM computing. SNNs are artificial neural networks that mimic natural neural networks and leverage timing of discrete spikes as the main information carrier. This section surveys the mainstream principles and algorithms of SNNs and their applications. 

\subsection{Bio-Inspired Computing Paradigms}
Bio-inspired photonic computing paradigms have emerged as a transformative approach that synergizes neurobiological principles with photonic device physics. These architectures exploit the inherent parallelism and ultrafast dynamics of light-matter interactions to overcome fundamental limitations in conventional electronic implementations. Recent advances demonstrate remarkable progress across multiple operational dimensions.

Photonic neural emulation leverages nonlinear optical phenomena to replicate biological neural dynamics. Silicon microring resonators with $8\times8$ weight banks implement leaky integrate-and-fire (LIF) models through intensity-dependent transmission characteristics, governed by:
\begin{equation}
    \tau\frac{dV}{dt} = -V + \sum_{i=1}^N w_iI_i - V_{th}\Theta(V-V_{th})
\end{equation}
where $\tau=15$ ps represents the membrane time constant, achieving 40 nm operational bandwidth with <0.5 dB/nm insertion loss \cite{Li2023NatPhoton}. Hybrid III-V/SOI platforms enable all-optical Hodgkin-Huxley neurons through carrier density modulation, demonstrating 20 GHz spiking frequencies that surpass biological counterparts by six orders of magnitude.

\begin{figure}[h]
\centering
\includegraphics[width=.8\textwidth]{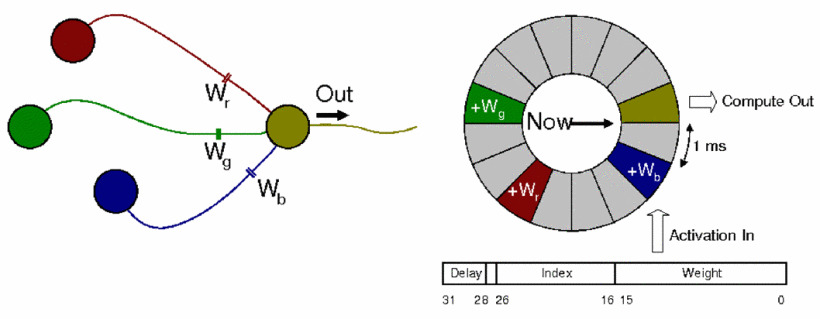}
\caption{SpiNNaker Neuron Binned Input Array. Inputs arrive into the bin
corresponding to their respective delay. Output is computed from the bin
pointed to by “Now”. 
\cite{}
}\label{fig:sixuan1}
\index{figures}
\end{figure}

Synaptic plasticity implementation has achieved unprecedented precision through material innovations. Phase-change Ge$_2$Sb$_2$Te$_5$ devices implement spike-timing-dependent plasticity (STDP) with:
\begin{equation}
    \Delta G = G_{\text{max}}\left[e^{-\Delta t/\tau_p} - e^{-\Delta t/\tau_d}\right]
\end{equation}
where $\tau_p=8$ ns and $\tau_d=12$ ns control potentiation/depression time constants, showing $10^6$ cycle endurance. Plasmonic Ag/TiO$_x$ memristors achieve 32 distinct conductance states with 5 aJ/spike energy efficiency, enabling reinforcement learning in photonic neural networks.

Neuromorphic sensory processing bridges biological fidelity with photonic precision. Graphene-CMOS hybrid photodetectors replicate retinal adaptivity through 120 dB dynamic range (0.1-10$^4$ lx) with 50 ps temporal resolution. Phononic crystal filters mimic cochlear frequency selectivity with 1/24 octave resolution (Q>500), while plasmon-enhanced waveguide arrays achieve parts-per-billion molecular detection limits for olfactory emulation.

Temporal information processing exploits photonic delay dynamics through nonlinear Schrödinger equation-governed interactions:
\begin{equation}
    i\frac{\partial A}{\partial z} = \frac{\beta_2}{2}\frac{\partial^2 A}{\partial T^2} - \gamma|A|^2A
\end{equation}
implementing STDP with 2 ps timing precision in dispersion-engineered fibers. Coupled microring resonators enable phase-encoded computation with <-150 dBc/Hz phase noise at 10 GHz, while low-loss Si$_3$N$_4$ waveguides (0.1 dB/cm) provide programmable delays up to 100 ns for working memory emulation.

\begin{table}[htbp]
\centering
\caption{Performance benchmarks of bio-inspired platforms}
\label{tab:benchmarks}
\begin{tabular}{lcccc}
\toprule
Technology & Latency (ps) & Energy/Spike & Bandwidth & Area Eff. \\
& & (aJ) & (THz) & (TOPS/mm$^2$) \\
\midrule
Electronic CMOS & 500 & 100 & 0.1 & 50 \\
Photonic STDP & 0.1 & 0.3 & 40 & 800 \\
Plasmonic RL & 1 & 5 & 200 & 1200 \\
Hybrid PCM & 10 & 5 & 10 & 400 \\
\bottomrule
\end{tabular}
\end{table}

The paradigm shift emerges from three fundamental advantages: (i) Photonic interconnects propagate signals at $0.64c$ with THz-class bandwidth, enabling 3D neural connectivity unattainable in metallic interconnects; (ii) Wavelength-division multiplexing supports $\geq$80 parallel channels in C+L bands, exponentially increasing network complexity; (iii) Optical nonlinearities (Kerr effect, $\chi^{(3)} \approx 10^{-14}$ m$^2$/W) provide activation functions without static power consumption.

Emerging frontiers focus on topological photonic neural networks for fault-tolerant computation and quantum-dot-based entangled state processing. Integration challenges are being addressed through heterogeneous III-V/SiN integration and inverse-designed meta-optics, potentially enabling exascale neuromorphic systems with <1 pJ/operation efficiency \cite{Zhang2024Optica}.

\begin{figure}[h]
\centering
\includegraphics[width=0.75\textwidth]{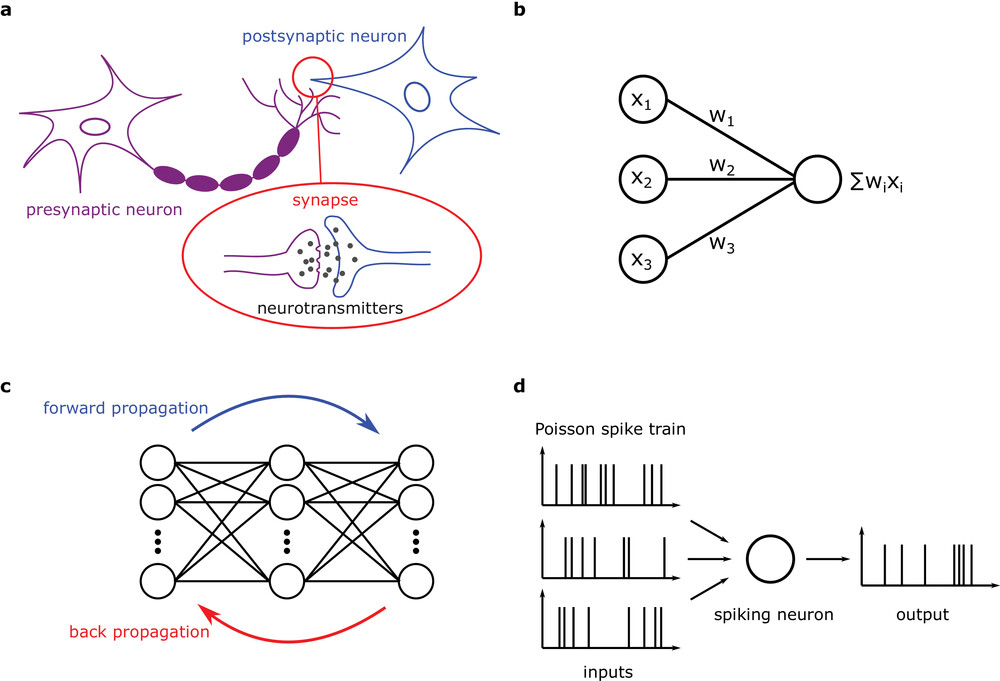}
\caption{SpiNNaker Neuron Binned Input Array. Inputs arrive into the bin
corresponding to their respective delay. Output is computed from the bin
pointed to by “Now”. 
\cite{}
}\label{fig:sixuan2}
\index{figures}
\end{figure}

\subsection{Spike Spatial-temporal Coding Schemes}
In the field of brain-inspired intelligence, efficient spike spatiotemporal coding schemes have always been a hot topic of research. They aim to provide a data representation form that is suitable for low-power, low-latency brain-like spike neural networks. Existing spike spatiotemporal coding schemes can be mainly divided into two categories: general coding and special coding.

\textbf{General Coding.} Neurons represent information primarily through two encoding methods. One is rate coding \cite{zhang1999geometrical, xu2024feel}, which focuses solely on the number of spikes within a unit of time while disregarding the specific timing of each spike. This limits its effectiveness in processing rapidly changing time-series information. The other is temporal coding \cite{theunissen1995temporal, nomura2022robustness, panzeri2001unified}, which takes into account both the number of spikes and the precise timing of each spike. This method excels in tasks requiring high temporal precision, as it can more accurately capture and process the complex dynamic relationships within time-series data. Currently, universal encoding has been successfully applied to tasks such as image classification, speech classification, gas classification, and multimodal recognition \cite{lin2025resistive, yao2025scaling}. However, universal encoding overlooks the unique attributes of each modality, leading to insufficient expression of key features and the generation of redundant spikes. This significantly degrades the performance of neuromorphic neural networks in multimodal tasks.

\textbf{Special Coding.} To enhance the adaptability of spike-timing encoding to various modalities, researchers have developed specialized encoding methods for specific modalities. In visual information encoding, Hopfield \cite{hopfield1995pattern} proposed a latency-based encoding scheme, which represents visual stimulus intensity through the spike timing of individual neurons. While effective for static visual information, it fails to encode dynamic visual information efficiently. To address this, Tobi Delbrück and colleagues \cite{gallego2020event} developed neuromorphic visual sensors, such as event cameras, which directly respond to light intensity changes with microsecond-level temporal resolution, achieving high sensitivity and speed. Based on this, Huang et al. \cite{huang20231000} developed a spiking camera that has been successfully applied to ultra-high-speed dynamic scenarios like image reconstruction and real-time wind turbine blade detection. In auditory information encoding, Pan et al. \cite{pan2020efficient} proposed a threshold-based encoding method, which monitors energy changes in speech signals using threshold neurons. However, this method generates excessive redundant information, reducing encoding efficiency and complicating subsequent neural network learning. To solve these issues, the team introduced a hybrid auditory encoding scheme that fully utilizes multi-dimensional information of speech signals, such as phase, frequency, and energy spectrum, achieving more comprehensive and efficient spiking speech encoding \cite{chen2023hybrid}. For olfactory information encoding, researchers designed an encoding strategy that maps gas information from the temporal domain to the frequency domain, converting gas concentration and composition into spatiotemporal spike signals for efficient representation \cite{borthakur2019spike}. Subsequently, Joon-Kyu et al. \cite{han2022artificial} proposed an integrated electronic nose sensor that replaces the olfactory organ, transmitting environmental odor information to a biological entity via electrical pulses, enabling high-intelligence human–computer interaction. In the tactile domain, researchers developed neuromorphic tactile sensors based on flexible electronic circuits, which measure force applied to the sensing area and represent this information through asynchronous event streams \cite{bartolozzi2018neuromorphic}. This has significantly advanced neuromorphic tactile tasks. Based on this, Bai et al. \cite{bai2023robotic} successfully developed a flexible robotic tactile perception system that can perform tasks such as texture recognition and pressure detection.

However, most of the aforementioned encoding methods are designed for single-task environments and cannot effectively adapt to complex and dynamic real-world scenarios, failing to ensure the effectiveness of spike encoding.

\subsection{Efficient models for SNNs}

Spiking Neural Networks (SNNs) integrate established deep learning architectures (e.g., CNNs and Transformers) with biologically inspired spiking neuron mechanisms to create efficient event-driven computing models. For example, MS-ResNet \cite{hu2021spiking} incorporates spiking neurons into residual networks \cite{he2016deep}, enhancing SNN depth and performance. TA-SNN \cite{yao2021temporal} combines attention mechanisms and spiking neurons to improve feature processing. Spikformer \cite{zhouspikformer} merges Transformers' self-attention with spiking neurons' discrete nature, using surrogate gradients to boost performance in tasks like depth estimation while reducing energy consumption. ANN-to-SNN conversion techniques (e.g., weight normalization, subtractive reset) enable mapping deep convolutional network parameters to spiking neurons, preserving performance and enabling sparse computation.

Taking inspiration from the brain's functional segmentation (e.g., visual and language areas), neuromorphic models adopt a modular design. This allows SNNs to be divided into functional modules (e.g., emergent and custom modules), mimicking the brain's parallel processing and dynamic coordination. For instance, the EB-NAS model \cite{pan2024brain}, inspired by multi-region brain structures and proposed by a team at the Chinese Academy of Sciences in 2025, uses neural circuit motifs to form dynamic functional modules and evolves cross-module connections. Configurable Foundation Model \cite{xiao2024configurable} spontaneously differentiates neuronal regions during pre-training (e.g., for math and code tasks), similar to the brain's lobes, and activates only relevant regions during inference for efficient computing. This design enhances model interpretability and supports distributed computing and continuous learning, offering new paths for general AI.

\begin{figure*}[htpb]
\centering\includegraphics[scale=0.5]{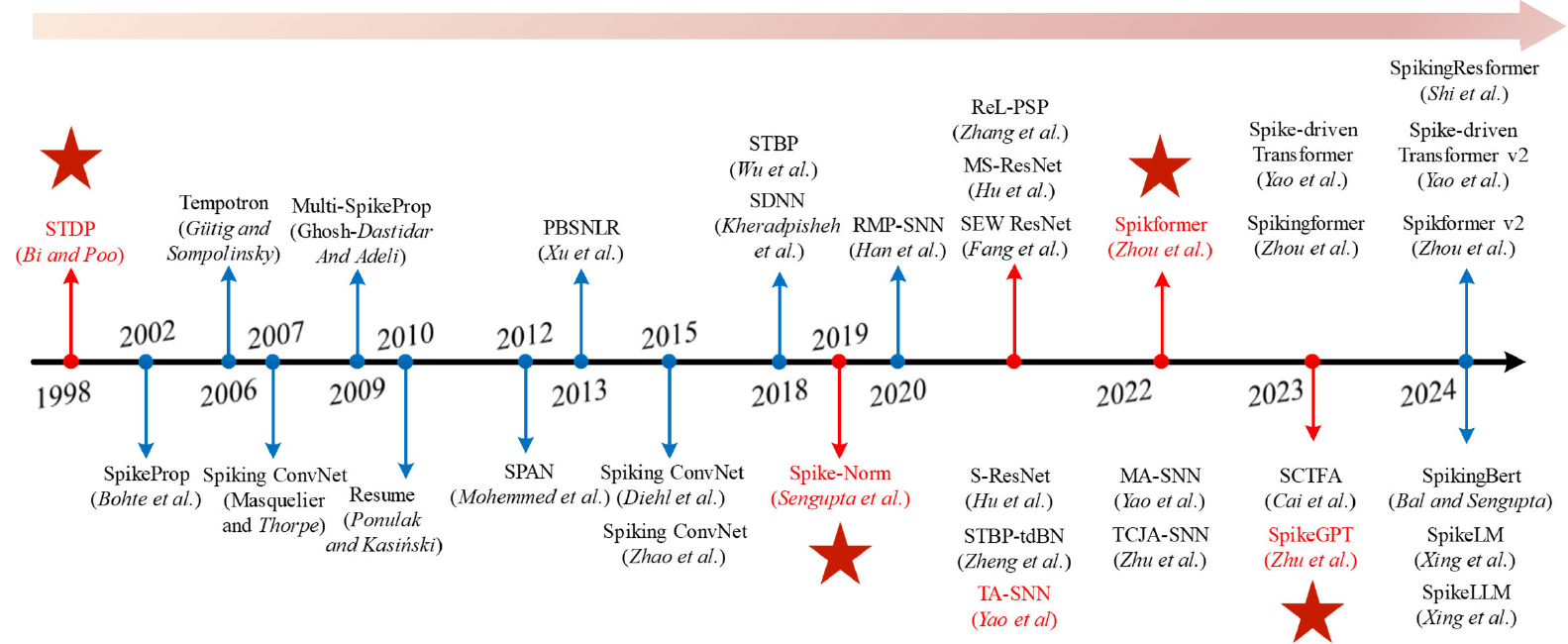}
\caption{ Roadmap of SNN: the evolution of spiking neural network architecture in five stages: feedforward, residual, attention-based, Transformer-based, and large language models. A red star marks the model seen as a significant landmark.}
\label{Fig:SNNs models}
\end{figure*}

Combining these two approaches creates a multimodal, deep spiking neural network model with the following features: It preserves the spatiotemporal coding and information transmission of biological neurons; it uses a hierarchical network structure to mimic the brain's spatial organization; and it enhances information representation and robustness through the collaborative working and information fusion of different functional brain-region networks. This design maintains biological plausibility while boosting network performance via multimodal coordination.

\subsection{Learning Algorithms}

Learning algorithms for brain-inspired SNNs primarily follow two research paths. One approach draws from neuroplasticity mechanisms to develop brain-inspired learning algorithms with biological interpretability. The alternative approach leverages optimization strategies from deep learning to develop high-performance SNN learning algorithms.

Brain-inspired learning algorithm development has explored biological plasticity mechanisms across various scales for different task types. Some works draw inspiration from microscale plasticity in biological systems, such as spike-timing-dependent plasticity (STDP) \cite{bi1999distributed}, to propose numerous unsupervised learning methods for single-layer SNNs \cite{guyonneau2005neurons, masquelier2007unsupervised}. However, these methods optimize network models only within local ranges, typically resulting in unstable convergence during network training, subsequently affecting model performance. To address these limitations, attention has shifted toward mesoscale plasticity mechanisms to enhance performance and training stability \cite{zhang2018plasticity, shi2020curiosity}. Further developments have incorporated macroscale plasticity principles to achieve global optimization from a holistic perspective \cite{zhang2021tuning, zhang2022multi}. Despite these advances, current brain-inspired SNN learning methodologies predominantly utilize neuroplasticity at a single scale, failing to effectively distribute spike credit across network outputs, hidden layer states, and local neural nodes in a comprehensive manner.

Recent research on high-performance algorithms inspired by deep learning has focused on integrating SNNs with deep neural networks, creating models that combine spike spatiotemporal representation capabilities with spatial hierarchical structures. Existing deep SNN learning methodologies can be categorized into two main approaches: ANN-to-SNN conversion methods and direct training methods. ANN-to-SNN conversion methods \cite{haobridging, huang2024towards, han2020rmp} initially train a traditional ANN model before transferring optimized synaptic weights to an SNN architecture, thus avoiding training challenges caused by non-differentiable discrete spikes while leveraging SNNs' low power consumption during inference; however, these methods disregard SNNs' temporal information processing capabilities and typically require higher inference latency to maintain conversion performance. In contrast, direct training methods \cite{wu2018spatio, meng2023towards, fang2023parallel}, employ surrogate gradient functions to address the non-differentiability issue of discrete spikes, thereby fully utilizing both the low power consumption advantages and temporal information processing capabilities of SNNs, achieving superior performance with reduced inference latency.

\subsection{Applications of Spiking Neural Networks}

With the increasing prominence of Spiking Neural Networks (SNNs) in low-power computing, their applications have extended to speech processing, visual understanding, and multimodal perception. In speech signal processing, a deep SNN model was proposed to address the challenge of capturing long-range dependencies in long audio sequences. By stacking one-dimensional dilated convolutions, the model effectively captures distant feature correlations while maintaining a compact parameter size, achieving improvements in both accuracy and energy efficiency, as reported in IEEE Transactions on Neural Networks and Learning Systems. In event-driven vision tasks, a deep SNN with a spatiotemporal attention mechanism was developed to enhance spatiotemporal feature extraction from event data, and was successfully applied to streaming object tracking and recognition with event cameras \cite{zhang2025spiking}. In auditory perception, a brain-inspired SNN-based sound source localization system was designed to address the poor robustness and low accuracy of conventional methods in complex environments. The system achieved state-of-the-art localization accuracy on public datasets, integrated neuromorphic hardware for ultra-low-power operation, and was successfully deployed on mobile robotic platforms \cite{pan2021multi}. Furthermore, to overcome the limitations of large model size and high energy consumption in conventional object detection and semantic segmentation models on edge devices, a spike-driven quantized SNN was proposed. This model achieved leading detection and segmentation performance on COCO and ADE20K datasets, reducing parameter count by 83.94\% and energy consumption by 79.36\%, while maintaining competitive accuracy compared to state-of-the-art convolutional neural networks \cite{qiu2025quantized}.

\begin{figure*}[htpb]
\centering\includegraphics[scale=0.47]{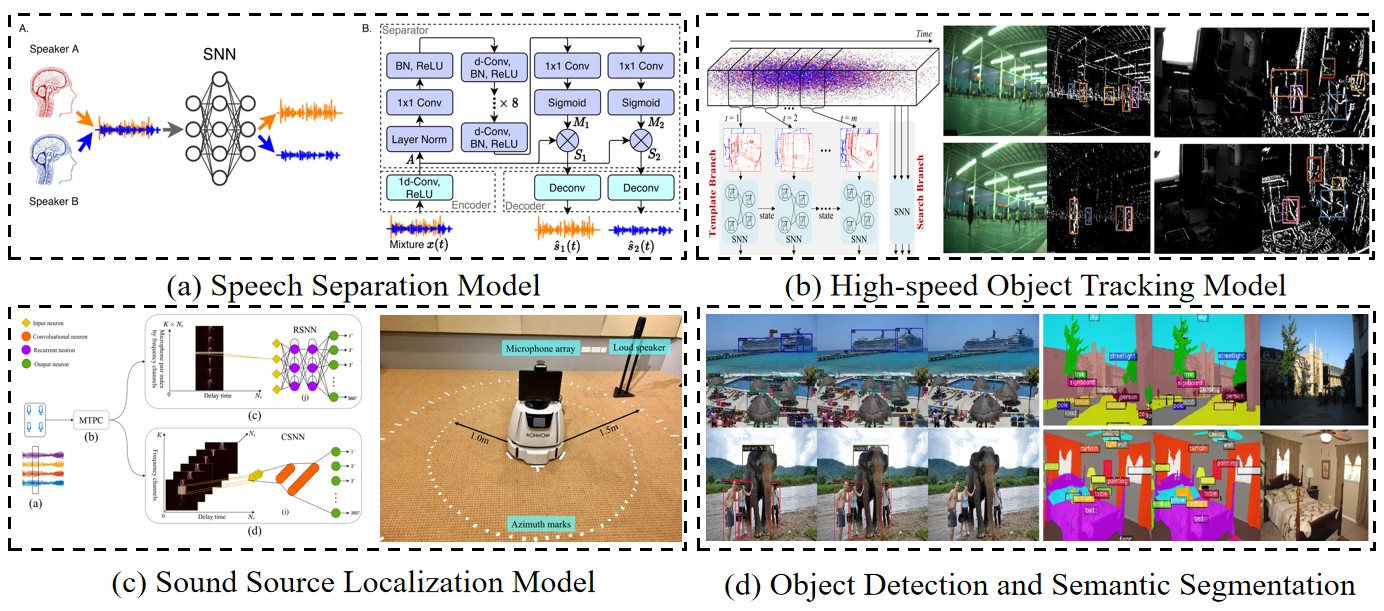}
\caption{Specific applications of SNNs: a. speech separation \cite{pan2021multi}, b. high-speed object tracking \cite{zhang2025spiking}, c. sound source localization \cite{pan2021multi}, d. object detection and semantic segmentation \cite{qiu2025quantized}.}
\label{Fig:applications of SNNs}
\end{figure*}

To support the efficient deployment of spiking neural networks (SNNs) in intelligent perception tasks such as sound source localization and image recognition, recent research has focused on improving energy efficiency and real-time performance in spike-based computation, aiming to design hardware architectures that are both lightweight and high-performing. Early studies developed SNN hardware systems \cite{ju2020fpga} that outperformed traditional CPUs and GPUs in inference speed on basic datasets such as MNIST, providing a basis for scaling to more complex tasks. The SIES computing engine proposed by the National University of Defense Technology \cite{wang2020sies} optimized neuronal dynamics through a unified pipelined array, significantly enhancing throughput and energy efficiency. The Skydiver architecture \cite{chen2022skydiver} introduced data partitioning and parallel processing to further improve system throughput under multi-input conditions. The DeepFire series \cite{aung2021deepfire} leveraged FPGA LUT resources by designing customized spike computing units, enabling efficient inference under resource constraints and supporting increasingly complex SNN models. FireFly and FireFly v2 \cite{li2023firefly, li2024firefly} utilized DSP48E2 units to build high-throughput ripple arrays, facilitating real-time inference for advanced SNNs across diverse tasks. Through the continuous advancement of hardware acceleration techniques, SNN deployment in tasks such as sound source localization and image recognition now achieves high energy efficiency, low latency, and multi-task processing capability, providing a strong foundation for intelligent perception systems in resource-constrained environments.

\section{Current Challenges and Future Directions}
\subsection{Memory issue with long context window and long token sequences}
Memory and Context Window: Photonic accelerators generally lack large on-chip memory to buffer long token sequences. Modern LLM inference may involve tens of thousands of tokens, requiring storage of activations, keys/values and intermediate states over the entire context. Without extensive SRAM or NVM on chip, photonic systems must stream these data in and out, reintroducing the von Neumann bottleneck. As Ning et al. observe, “data movement frequently constitutes the bottleneck of the entire system” – a problem that applies “not only in traditional electronic processors but also in optical processors”. In practice, limited on-chip memory forces a photonic LLM implementation to fetch context from external DRAM or disks, incurring latency and breaking the all-optical pipeline. Emerging use cases like retrieval-augmented generation exacerbate this: performing near-real-time search and tokenization of multi-terabyte text corpora adds another round of expensive memory access. In short, the finite storage capacity of photonic chips constrains the feasible context length and throughput for LLMs, making long-sequence inference a major challenge.

\subsection{Storage issue with mega-sized datasets on photonic computing systems}
Storage and I/O Bottlenecks: Large-language models and their training or knowledge bases involve enormous datasets (multiple terabytes). Photonic accelerators still depend on high-bandwidth external memory and storage to feed these data. The I/O bandwidth needed can easily outstrip the available interfaces: even if the optical core is extremely fast, it is wasted if data cannot be streamed in quickly enough. Analysts warn of a growing “memory wall” for LLMs, where moving data becomes the dominant limitation. This is compounded by real-world workloads: for example, retrieval-augmented LLMs must repeatedly fetch and process large text blocks, placing severe demands on I/O. Some proposals (like co-locating non-volatile weight storage) can cut I/O (one study reports a 1000× reduction in chip I/O by using on-chip flash for weights), but even so the scale of multi-terabyte corpora means that data staging, caching, and bus bandwidth will remain critical bottlenecks in photonic LLM systems.

\subsection{Precision and Conversion Overhead}
Photonic computing is intrinsically analog, so representing high-precision tensors (needed for LLM inference) is difficult. State-of-the-art photonic Transformer designs rely on high-resolution ADCs/DACs to preserve accuracy, and these converters consume the majority of chip area and power. For instance, in one photonic transformer accelerator the ADC/DAC circuitry occupied over 50\% of the chip and became a performance bottleneck. Reducing quantization error without blowing up conversion overhead is an ongoing challenge: low-bit converters or shared ADC schemes can improve area/energy, but may hurt model fidelity. Thus, finding optimal analog quantization schemes or mixed-signal architectures (perhaps using digital correction for a small fraction of values as in) is critical for next-generation photonic LLM chips.

\subsection{Lack of Native Nonlinear Functions}
Finally, photonic hardware excels at linear operations (matrix-vector multiplies via interferometers) but historically has lacked easy ways to implement activation and nonlinear layers. Early integrated photonic neural networks could perform fast matrix multiplies but needed electronic circuits for activation functions. In practice, many proposed photonic LLM accelerators still require conversions to CMOS for softmax, GELU, and other pointwise functions. Integrating efficient on-chip nonlinear elements (e.g. optical saturable absorbers, electro-optic modulators, or nanophotonic nonlinearities) or developing hybrid optical-electrical pipelines that minimize these gaps is a key engineering hurdle for fully optical LLM inference

\subsection{Photonic Attention Architectures}
A major research thrust is to implement transformer self-attention directly in optics. This involves designing tunable photonic weight elements and reconfigurable interferometer networks to compute QxK and V-weighted sums optically. For example, photonic tensor cores are being developed that use Mach–Zehnder interferometer (MZI) meshes or other crossbar arrays to carry out large matrix multiplications in parallel. Tunable weights may be realized by phase shifters, microring modulators, or even magneto-optic memory cells: one recent proposal used Ce:YIG resonators to store multibit weights, enabling non-volatile, on-chip optical weight storage. In addition, delay-based schemes from reservoir computing could provide temporal context: long optical delay lines or series-coupled microrings have demonstrated very high memory capacity for sequential tasks. A promising vision is an all-optical transformer block where dynamic weight matrices are programmed into an optical mesh and past token states are held in transit delays, allowing the self-attention kernel to be evaluated at light speed. Recent designs like Lightening-Transformer (a “dynamically-operated photonic tensor core”) and HyAtten validate this approach: they achieve highly parallel, full-range matrix operations while minimizing off-chip conversion. Continued work on integrated optical buffers, high-bandwidth modulators, and photonic softmax approximations will advance this direction.

\subsection{Neuromorphic and Spiking Photonic LLMs}
Another pathway is to recast LLM inference in a neuromorphic, event-driven paradigm. Spiking neural networks (SNNs) process data as sparse asynchronous events, which naturally match photonics’ strengths. Indeed, all-optical spiking neural networks have been demonstrated on chip using phase-change neurons and laser pulses. One could imagine encoding a token stream as optical spikes or pulses and using a photonic SNN with synaptic weights to perform sequence processing. Hybrid photonic–spintronic designs could play a role here: spintronic devices (magnetic tunnel junctions, phase-change synapses) provide compact non-volatile weight storage and can interface with optical neurons. Recent work on photonic in-memory weights (using magneto-optics) and on photonic neuromorphic accelerators leveraging extreme sparsity suggests that embedding non-linear, event-driven components on a photonic chip is feasible. Such architectures could exploit data sparsity (most tokens only weakly excite the network) and update weights only when events occur, greatly reducing energy. Exploring spiking attention models or sparse transformer variants on photonic neuromorphic hardware is an exciting future direction for low-power LLM inference.

\subsection{System Integration and Co-Design}
Finally, scaling LLMs on photonics will require co-design across layers. This includes integrating photonic processors with advanced optical I/O and memory hierarchies, as well as co-optimizing algorithms for the hardware’s strengths. For example, recent fully integrated photonic DNN chips (fabricated in commercial foundries) show it is possible to perform all neural network computations optically on-chip. Extending such integration to transformer-scale models will demand dense wavelength-division multiplexing, optical network-on-chip fabrics, and novel packaging (e.g. co-packaged optics) to boost throughput. Meanwhile, software tooling (quantization, parallelism, placement) must adapt to photonic hardware. Efforts on photonic-electronic co-packaging and compute-in-memory architectures offer a roadmap: by tightly coupling photonic tensor cores with co-located memory banks and control logic, one can mitigate the von Neumann overhead. In the longer term, success will likely come from global co-design — matching transformer algorithms (sparsity, low precision, model partitioning) to the capabilities of non-von Neumann photonic chips. These combined hardware/software innovations could unlock the massive parallelism of light for next-generation LLM workloads.

\section{Conclusion}
Advances in photonics have catalyzed a transformation in computational technologies, with the integration of optoelectronics onto photonic platforms leading the charge. This integration has facilitated the emergence of PICs, which act as the building blocks for ultra-fast artificial neural networks and are pivotal in the creation of next-generation computational devices. These devices are engineered to address the intensive computational demands of machine learning and AI applications across sectors including healthcare diagnostics, complex language processing, telecommunications, high-performance computing, and immersive virtual environments.

Despite the advancements, conventional electronic systems exhibit limitations in speed, signal interference, and energy efficiency. Neuromorphic photonics, characterized by its ultra-low latency, emerges as a groundbreaking solution, carving out a new trajectory for the advancement of AI and ONNs. This review casts a spotlight on the latest developments in neuromorphic photonic systems from the perspective of photonic engineering and material science, critically analyzing the emergent and anticipated challenges, and mapping out the scientific and technological innovations necessary to surmount these obstacles.

The focus is on an array of neuromorphic photonic AI accelerators, examining the spectrum from classical optics to sophisticated PIC designs. It scrutinizes their operational efficiency, particularly in terms of operations per watt, through a detailed comparative analysis emphasizing key technical parameters. The discussion pivots to specialized technologies such as VCSEL/PCSEL and frequency microcomb-based accelerators, accentuating the latest innovations in photonic modulation and wavelength division multiplexing for effective neural network training and inference.

Acknowledging the current technological barriers in achieving computational efficiencies at the PetaOPs/Watt threshold, the review explores prospective strategies to enhance these critical performance metrics. These include the emerging topological insulators and PCSELs as well as strategies to advance fabrication, system scalability and reliability. The exploration aims to not only chart the current landscape but also to forecast the trajectory of neuromorphic photonics in pushing the frontiers of AI capabilities in the near future. All in all, as the Moore's law comes to an end and the photonic version of the Moore's law begins to take off, we expect to see a considerable improvement in PIC's cost, scalability, integratability, and total computing capacity. PICs will eventually replace ICs as the backbone of future computing systems.

\section*{Acknowledgement}
This work is supported by National Natural Science Foundation of China (NSFC) under Grant No.62174144, Shenzhen Science and Technology Program under Grant No.JCYJ20210324115605016, No.JCYJ20210324120204011, No.JSGG20210802153540017, No.KJZD20230923115114027, and No.JCYJ20220818102214030, Guangdong Key Laboratory of Optoelectronic Materials and Chips under Grant No.2022KSYS014, Shenzhen Key Laboratory Project under Grant No.ZDSYS201603311644527; Longgang Key Laboratory Project under Grant No.ZSYS2017003 and No.LGKCZSYS2018000015; Shenzhen Research Institute of Big Data; Innovation Program for Quantum Science and Technology under Grant No.2021ZD0300701.\par

\section*{Conflict of interests}
The authors declare no conflict of interests.

\section*{Author contributions}
Z.Z conceived and proposed the writing project. R.L. and Z.Z. designed the structure and outline of the paper. R.L. led and arranged the writing of the whole paper. R.L., Y.G., H.H., Y.Z., S.M. wrote the paper together. S.M. and Y.Z. applied for copyright permissions. Z.Z. supervised and mentored the project. Z.Z. funded the project.

\renewcommand*{\bibfont}{\normalfont\small}

\printbibliography

\end{document}